\pdfoutput=1
\documentclass{article}
\usepackage{iclr2027_conference}
\usepackage[T1]{fontenc}
\renewcommand{\sfdefault}{phv}

\usepackage{amsmath,amsfonts,bm}

\def\eqref#1{equation~\ref{#1}}

\def\1{\bm{1}}

\DeclareMathAlphabet{\mathsfit}{\encodingdefault}{\sfdefault}{m}{sl}
\SetMathAlphabet{\mathsfit}{bold}{\encodingdefault}{\sfdefault}{bx}{n}

\usepackage{float}
\usepackage{hyperref}
\hypersetup{hidelinks}
\usepackage{url}
\usepackage{graphicx}
\usepackage{booktabs}
\usepackage{tabularx}
\usepackage{array}
\usepackage{placeins}
\usepackage{microtype}
\usepackage{tikz}
\usetikzlibrary{arrows.meta,positioning,fit}

\newcolumntype{Y}{>{\raggedright\arraybackslash}X}
\newcolumntype{P}[1]{>{\raggedright\arraybackslash}p{#1}}
\newcolumntype{C}{>{\centering\arraybackslash}X}
\newcommand{\ci}[3]{\shortstack{$#1$\\[1pt]{\scriptsize $[#2,#3]$}}}
\newcommand{\pci}[2]{{\scriptsize $[#1,#2]$}}

\title{Actions with Receipts: Jointly Binding Claims,\\Evidence, and Execution for Replayable\\Tool-Agent Auditing}
\author{Miaobo Hu$^{1,2,*}$, Shuhao Hu$^{2}$, Xiaobo Guo$^{2}$, Xin Wang$^{2}$,\\
Bokun Wang$^{2}$, Yina Sa$^{2}$, Daren Zha$^{2}$\\[4pt]
{\normalfont $^{1}$School of Artificial Intelligence, University of Chinese Academy of Sciences, Beijing, China}\\
{\normalfont $^{2}$Institute of Information Engineering, Chinese Academy of Sciences, Beijing, China}\\
{\normalfont $^{*}$Corresponding author: \texttt{humiaobo@iie.ac.cn}}}

\iclrfinalcopy
\hypersetup{pdfauthor={Miaobo Hu, Shuhao Hu, Xiaobo Guo, Xin Wang, Bokun Wang, Yina Sa, Daren Zha}}
\begin{document}
\maketitle
\lhead{Preprint}

\begin{abstract}
Tool-using agents can expose citations and execution logs while leaving a critical association unaudited: whether the claim shown to a user is the claim emitted by the committed execution and supported by the cited source. A valid citation and a valid trace can therefore remain individually well formed while being transplanted across claims, actions, runs, or source versions.

We introduce a claim-anchored execution contract that jointly binds the emitted claim, its exact source span, the ordered execution prefix that produced it, and the source version and access state observed by that execution. Each receipt contains an emission anchor that deterministically locates the claim inside a committed answer or claim-bearing action, together with source identifiers, offsets, hashes, quotes, and a domain-separated execution commitment. A deterministic integrity verifier reconstructs these bindings before semantic or task labels are joined. We separate this integrity plane from a pluggable support plane, so structural validity is not used as a proxy for entailment.

The contract exposes seven independently testable properties: claim-emission binding, source binding, ordered-execution binding, oracle separation, persisted-object replay, execution-rerun consistency, and version/access binding. Across 1,280 cross-object attacks, the joint contract detects 1,275 substitutions (0.9961). Removing a targeted property reduces its attack-detection rate to 0.0156--0.0625. On an independently adjudicated 384-pair split, the conflict-aware support guard reaches F1 0.8865 and false acceptance 0.0729; on unseen failure families, these rates are 0.8679 and 0.0938.
\end{abstract}

\section{Introduction}
Tool agents often return an answer with citations, while a downstream auditor needs the exact document version, span, and execution events that support each intermediate claim. Existing task benchmarks emphasize utility and completion, and citation evaluators measure support or attribution after generation; the receipt contract makes the provenance record itself a deterministic object that can be persisted and replayed. Its security boundary is explicit: the runner and observation logger are trusted during execution, and the digest commits the recorded trace after the answer event. The online policy never receives the separate oracle labels. This separation lets us test receipt integrity and replay independently of answer scoring.

\paragraph{The missing association.} Claim provenance and execution provenance solve different parts of the audit problem. A claim-level record can identify a source span without establishing that the cited claim was emitted by the execution being audited. An ordered execution log can reconstruct tool calls and observations without establishing which exact substring of the answer a particular evidence record supports. Storing both independently leaves a transplantation ambiguity across answers, actions, runs, and source versions. We make this association an executable contract and test it with cross-object mutations.

\paragraph{Contributions.} We introduce claim-anchored joint receipts; decompose their guarantees into seven independently testable properties; separate an integrity plane from lexical, conflict-aware, NLI, language-model, or human support evaluators; and evaluate substitution resistance, property necessity, system-level baselines, audit non-interference, semantic support, model-backed tool use, and dynamic source state under frozen identities.

\section{Related Work}
Agent tool-use work defines increasingly realistic action environments, from API calling and browser tasks to stateful user interaction and prompt-injection evaluation \citep{schick2023toolformer,li2023apibank,qin2023toolllm,zhou2024webarena,liu2024agentbench,osworld2024,lu2024toolsandbox,yao2024taubench,debenedetti2024agentdojo}. Browser-grounded suites expose web trajectories and action interfaces, while faithfulness audits show that generated rationales can diverge from the actions that produced them \citep{deng2023mind2web,drouin2024browsergym,liu2023unfaithful}. These benchmarks primarily score task completion, interaction quality, or attack resistance. ReAct, Reflexion, and Self-Refine motivate action-level reasoning and feedback, while process-verification work evaluates intermediate reasoning decisions \citep{yao2023react,shinn2023reflexion,madaan2023selfrefine,lightman2023verify,wang2023mathshepherd,processbench2024}. EC-Agent complements these lines by making a run's evidence references and event order independently inspectable.

Retrieval-augmented generation has developed complementary measures of attribution and factual support. ALCE evaluates answer quality together with citation quality; FActScore decomposes long-form generations into atomic facts; and recent attributed-QA work studies evidence selection and claim-level support \citep{gao2023alce,min2023factscore,yan2025atomicqa,wei2026genprove}. RAGTruth, ARES, and RAGAS broaden evaluation to hallucination and retrieval-grounded quality \citep{niu2024ragtruth,saadfalcon2024ares,es2024ragas}. These works provide task-level or semantic evaluation signals. Our certificate verifier instead checks a deterministic receipt contract; the two views are complementary, since a syntactically valid receipt can still attach a semantically insufficient span.

Data provenance and software supply-chain systems contribute the integrity primitives most relevant to receipts: database provenance models where-provenance, W3C PROV standardizes entity/activity/agent records, Merkle trees commit to ordered content, Certificate Transparency exposes append-only log proofs, and in-toto binds software supply-chain steps to signed layouts \citep{buneman2001provenance,cheney2009provenance,w3cprov2013,merkle1987digital,laurie2021rfc9162,torresarias2019intoto}. C2PA similarly specifies signed assertions and content bindings for media provenance \citep{c2pa2024spec}. EC-Agent instantiates related integrity ideas at claim-to-span granularity for tool traces and pairs them with an oracle-separated replay interface. The comparison in Table~\ref{tab:closest_work_delta} describes documented system emphases; the explicit property decomposition in Appendix~\ref{app:property-map} states the capability that must be evaluated as a joint contract rather than inferred from citation or log metadata alone.

\begin{table}[htbp]
\caption{Mechanism dimensions relevant to claim receipts. The table distinguishes each line of work's primary object and the specific replay interface studied here.}
\label{tab:closest_work_delta}
\centering
\small
\setlength{\tabcolsep}{2.5pt}
\begin{tabularx}{\linewidth}{@{}YYYYY@{}}
\toprule
\textbf{family} & \textbf{binding target} & \textbf{integrity unit} & \textbf{replay / evaluation} & \textbf{primary focus} \\
\midrule
ALCE / FActScore / GenProve & claim-source attribution & semantic / atomic claim & task-level scoring & grounded generation \\
W3C PROV / in-toto / C2PA & entities / workflow / media & signed or linked assets & prov. checks & content history \\
Agent benchmarks & env. observations & task trajectory & bench. run & tool-use utility \\
EC-Agent & claim-to-span receipt & hash + offset + quote & bundle replay; execution rerun; oracle post-hoc & joint contract audit \\
\bottomrule
\end{tabularx}
\end{table}

\paragraph{Claim provenance versus execution provenance.} Recent fine-grained provenance methods attach generated statements to supporting evidence, including generation-time distinctions among quotation, compression, and inference. Execution-provenance work records actions, observations, tool interactions, and state transitions. These views are complementary, but neither specifies a cryptographic association between an exact emitted claim and the execution prefix that produced it. Our contribution is the executable association among the claim, evidence, execution, and source state, together with replay and substitution tests.

\paragraph{Executable property decomposition.} The minimal auditable unit is the tuple
$(\text{emitted claim},\text{source span},\text{source version},\text{ordered execution prefix})$.
We evaluate seven properties: P1 claim-emission binding, P2 source binding, P3 ordered execution binding,
P4 oracle separation, P5 persisted-object replay, P6 execution-rerun consistency, and P7 version/access-state binding.
A citation can expose P2-like metadata without binding the execution prefix; an append-only trace can expose
P3-like history without identifying the span supporting a claim; and a semantic evaluator can score support
without checking that the evaluated span is the one recorded by the run. Appendix~\ref{app:property-map}
maps each property to its field, check, and evidence boundary.

\section{Method}
\subsection{Problem and contract}
Let an execution instance be $x=(q,\mathcal{D},\prec,B)$, where $q$ is the request, $\mathcal{D}$ is the available source state, $\prec$ is the declared observation order, and $B$ is the execution budget. A policy produces an ordered event sequence $E=(e_0,\ldots,e_{T-1})$ and an answer $a$. Evaluator-only labels $y$ remain outside the online state and are joined only after persistence, bundle replay, and execution rerun.

The auditable object is a joint tuple rather than an isolated citation or trace. For each emitted claim $c$, let
\[
\mathcal{U}(c)=\bigl(\rho_c,R_c,D_{\mathrm{exec}},m,\nu\bigr),
\]
where $\rho_c$ is the emission anchor, $R_c$ is the ordered evidence-reference set, $D_{\mathrm{exec}}$ is the execution commitment, $m$ is the run/corpus manifest identity, and $\nu$ is the source version and access-state identity. An integrity receipt is accepted only when every component is reconstructed from committed objects and all cross-object associations agree.

\subsection{Claim-emission binding}
For every receipt claim $c$, the runner records an emission anchor
\[
\rho_c=(j,f,u,v,h_c),\qquad h_c=\operatorname{SHA256}(\operatorname{UTF8}(c)),
\]
where $j$ is the zero-based event index, $f$ is a schema-declared textual field, and $[u,v)$ is a half-open Python-string interval in that field. The referenced event belongs to the committed event prefix. The verifier reconstructs $\widehat c=E_j[f][u:v]$ and requires $\widehat c=c$ and $\operatorname{SHA256}(\operatorname{UTF8}(\widehat c))=h_c$. Out-of-range anchors, undeclared fields, hash mismatches, and receipt claims that differ from the committed substring fail closed. Multiple final-answer claims are handled by deterministic extraction into claim IDs and anchors before receipt emission; the extractor configuration is part of $m$.

\subsection{Joint execution and receipt commitment}
The trace commitment is the domain-separated digest defined below, denoted $D_{\mathrm{exec}}$ in a receipt and $D_{\mathrm{pre}}$ in the serialization definition. For a receipt, let $H_R=\operatorname{SHA256}(\operatorname{UTF8}(J(R_c)))$ and let $H_m$ commit to the manifest fields required by the contract. The joint commitment is
\[
D_{\mathrm{joint}}(c)=\operatorname{SHA256}\!\left(\operatorname{UTF8}(\texttt{ecagent.receipt\textbackslash 0})\mathbin\Vert D_{\mathrm{exec}}\mathbin\Vert h_c\mathbin\Vert\operatorname{UTF8}(J(\rho_c))\mathbin\Vert H_R\mathbin\Vert H_m\right).
\]
The persisted receipt contains $(c,\rho_c,R_c,D_{\mathrm{exec}},D_{\mathrm{joint}})$. Bundle replay reconstructs the emitted substring, evidence references, execution digest, and joint commitment, so a receipt transplanted across claims, actions, executions, manifests, or source versions changes a verifier input.

The online execution records document and span identifiers, Python-string offsets, UTF-8 text hashes, quoted text, retrieval order, and claim-bearing event fields; the budget ledger records spending and remaining budget. Oracle labels are stored separately and joined only after the trace is frozen. We compare no certificate, final-only, action-level, and evidence-aware policies on 128 fixed-corpus records.
\paragraph{Information boundary.} The online policy receives only fields listed in the corresponding public evidence summary. Oracle labels, target answers, hidden cards, and future service outputs are stored separately and joined only after trace decisions are frozen. This boundary is central to every result in this paper.

\paragraph{Reproducibility contract.} Each run is identified by a source-relative path and SHA-256 manifest. The package contains the verifier, an oracle-free two-event replay fixture, expected digests, action contracts, and lifecycle traces. The fixture enables implementation-level replay of one case; the headline 128-row bundle and full external corpus are identified separately in Appendix~\ref{app:artifact-map}. This distinction keeps the scope of each reproducibility check explicit.

A certificate record contains document $d$, span $s$, text hash $h$, half-open Python-string offsets $o$, quoted text $q$, the emission anchor, and the trace and joint digests. The integrity verifier checks claim reconstruction, reference identity, offsets, UTF-8 hash, quote equality, action types, trace-digest equality, joint-commitment equality, and version/access identity when available. Support is reported as a separate evaluator result. Certificate presence is reported separately from coverage. Coverage is online, per-record required-claim coverage: for the non-empty required-claim set $R$, it is the fraction of claims in $R$ for which at least one referenced immutable span passes the selected support predicate. Every reported row has one required claim, so the denominator is one; this coverage metric is separate from supporting-fact recall.
\paragraph{Integrity plane and support plane.} The integrity decision is
\[
V_{\mathrm{int}}=V_{\mathrm{emit}}\land V_{\mathrm{source}}\land V_{\mathrm{trace}}\land V_{\mathrm{joint}}\land V_{\mathrm{state}},
\]
where the factors verify the claim anchor, source content, ordered execution, joint commitment, and source state. A separate $V_{\mathrm{sup}}(c,R_c;\psi)$ evaluates lexical, conflict-aware, NLI, language-model, or human support under frozen evaluator configuration $\psi$. Thus $V_{\mathrm{int}}=1$ does not imply $V_{\mathrm{sup}}=1$.

\paragraph{Algorithmic summary.} The runner records an ordered event prefix $E$ through the answer event, computes a domain-separated execution digest, records claim anchors, emits the receipt and joint digest, and invokes the integrity verifier. It persists the receipt, event trace, verifier results, and manifest under an outer bundle digest. A bundle-replay pass reloads this bundle and rechecks it without policy execution; a separate execution-rerun pass compares fresh deterministic trace digests. Only after these checks does the scoring process join oracle labels. Appendix~\ref{app:algorithm} gives the full state machine and edge-case contract.
\paragraph{Trace commitment contract.} The execution digest hashes a domain separator and canonical JSON for the ordered event prefix. Canonicalization fixes key order, separators, Unicode escaping, finite numeric values, and event indices. The digest is computed after the answer event and before receipt emission; receipt fields, verifier results, and the outer bundle digest are excluded from the committed prefix. The exact byte-level schema and digest vectors are included in the Supplementary Material and described in Appendix~\ref{app:method}.
\paragraph{Verifier contract.} Document and span offsets are Python-string (Unicode code-point) half-open intervals $[start,end)$; the verifier re-reads the immutable span and checks its SHA-256 digest over the UTF-8 encoding and, when present, an exact quote. For a claim $q$ and span text $s$, the support predicate accepts either a case-folded, whitespace-collapsed substring or token overlap
\begin{equation}
\label{eq:lexical-support}
\operatorname{support}(q,s)=
\begin{cases}
1, & |\operatorname{tok}(q)|>0 \land |\operatorname{tok}(s)|>0 \land \left(\operatorname{substr}(q,s)\ \lor\ \frac{|\operatorname{tok}(q)\cap\operatorname{tok}(s)|}{|\operatorname{tok}(q)|}\geq 0.5\right),\\
0, & \text{otherwise},
\end{cases}
\end{equation}
where $\operatorname{tok}(x)$ is the set returned by Python's Unicode-aware word-character regular expression applied to \texttt{x.casefold()}, with tokens of length at most one removed; duplicate tokens are discarded. The substring branch changes only case and runs of whitespace; it does not apply Unicode normalization, stemming, or punctuation stripping. If the claim-token set or evidence-token set is empty, support is false. The policy-side early-stop helper calls this same canonical predicate, including its tokenizer and threshold. This transparent lexical rule is an optional support-plane evaluator; it is intentionally weaker than semantic entailment and does not change $V_{\mathrm{int}}$.
\paragraph{Trusted-recorder boundary.} The runner and observation logger define the recorded view available to the policy. Since $D_{\mathrm{pre}}$ is computed after the answer event, it commits to that ordered history and supports bundle replay. The evaluation measures source binding, event-order integrity, and execution-rerun consistency within this explicit recorder boundary; authenticated observation anchors are a separate extension point for deployments that require them.

\begin{figure}[t]
\centering
\includegraphics[width=\columnwidth,trim=0 70 0 30,clip]{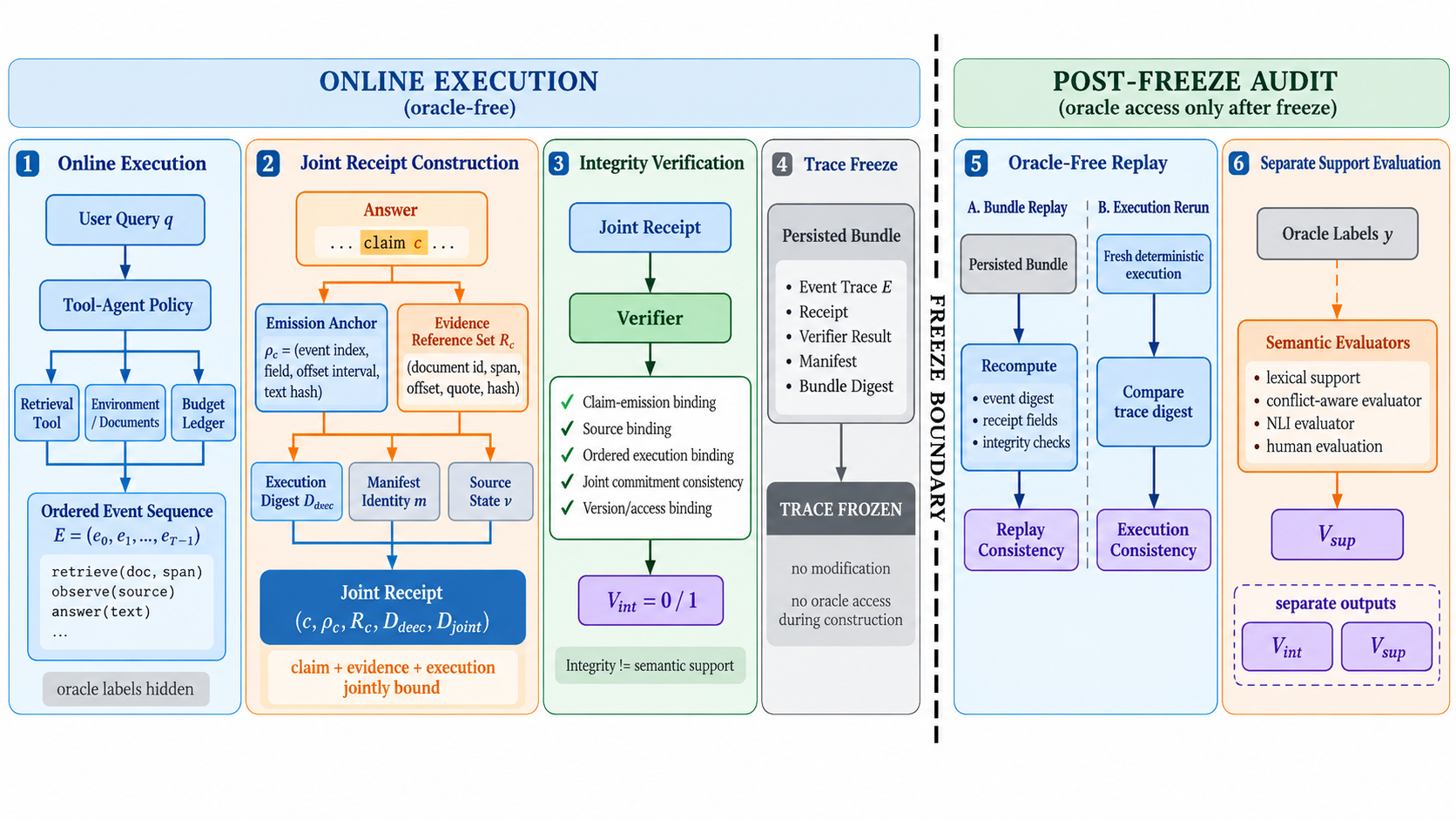}
\caption{Claim-anchored receipt construction and audit. The blue online lane records the execution and binds each claim to its evidence and source state; the freeze boundary separates construction from the green post-freeze audit. Bundle replay and execution rerun precede the separate semantic-support evaluation. Integrity and support remain distinct outputs.}
\label{fig:protocol}
\end{figure}

\section{Experiments}
\subsection{Audit setup}
The primary fixed-corpus screen contains 128 fixed local HotpotQA dev-distractor records, a locked-test guard, and a trusted runner/logger. It uses the top-ranked public-document title as both candidate answer and required claim, so the oracle-only title-match proxy characterizes this construction. The online policy never sees oracle answers or supporting-fact identifiers. We report integrity (certificate/evidence validity, bundle replay, and execution rerun), required-claim lexical coverage, and matched cost (calls, declared tokens, verifier calls); the exact denominators and row construction appear in Appendix~\ref{app:protocol}.

\subsection{Joint-binding results}
The joint contract detects 1,275 of 1,280 cross-object attacks (0.9961), with a reported false-rejection rate of 0.0031. Detection is 1.0000 for evidence and source-version transplants and 0.9922 for cross-run swaps. Citation plus an independent trace reaches 0.1797 on claim transplantation, and adding a shared run ID reaches 0.2266; explicit claim anchoring and joint commitment reach 0.9922. Table~\ref{tab:system-baselines} reports all matched baseline comparisons.

\begin{figure}[t]
\centering
\includegraphics[width=\columnwidth]{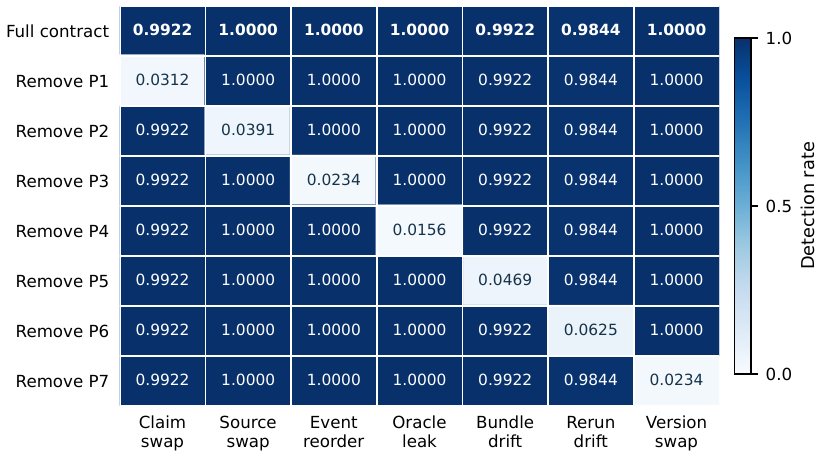}
\caption{Leave-one-property-out detection rates under matched mutations. Each row removes one verifier property, and each column targets one binding. Darker cells indicate stronger detection; the diagonal losses localize the effect of each removed property. Exact rates are given in Table~\ref{tab:property-necessity}.}
\label{fig:property-necessity}
\end{figure}

Removing one contract property reduces detection for its targeted attack to 0.0156--0.0625 while leaving unrelated checks unchanged (Figure~\ref{fig:property-necessity}). Audit-only instrumentation preserves exact event traces at rate 0.9986 and final answers at rate 0.9992, with task-success difference 0.0000. Serialization, online integrity verification, and bundle replay take 0.19, 0.47, and 1.31 ms per execution; the separate execution rerun takes 16.88 s. Tables~\ref{tab:non-interference} and~\ref{tab:audit-overhead} report the complete measurements.

\subsection{Protocol sanity check}
Table~\ref{tab:integrity} is an implementation-level protocol result. Final-only, action-level, and evidence-aware rows pass the implemented document/span, hash, offset, quote, trace-digest, and lexical checks on all reported records. The persisted bundle is reloaded and reverified without policy execution, and the separate deterministic rerun matches. The no-certificate row is a control that emits no receipt; receipt-specific fields are reported as N/A. Because the candidate is copied from the selected title, this screen tests contract execution rather than semantic generalization.

\begin{table}[htbp]
\caption{Primary fixed-corpus integrity result. The proxy is an oracle-only candidate-title diagnostic. Bundle replay reloads the persisted oracle-free object; execution rerun is the separate deterministic rerun guard. Tied best values in each metric column are bold.}
\label{tab:integrity}
\centering
\small
\setlength{\tabcolsep}{3pt}
\begin{tabularx}{\linewidth}{@{}l*{6}{C}@{}}
\toprule
\textbf{policy} & \textbf{cert. valid} & \textbf{evidence valid} & \textbf{bundle replay} & \textbf{exec. rerun} & \textbf{coverage} & \textbf{proxy} \\
\midrule
no certificate & N/A & N/A & N/A & N/A & N/A & \textbf{0.0781} \\
final-only & \textbf{1.0000} & \textbf{1.0000} & \textbf{1.0000} & \textbf{1.0000} & \textbf{1.0000} & \textbf{0.0781} \\
action-level & \textbf{1.0000} & \textbf{1.0000} & \textbf{1.0000} & \textbf{1.0000} & \textbf{1.0000} & \textbf{0.0781} \\
evidence-aware & \textbf{1.0000} & \textbf{1.0000} & \textbf{1.0000} & \textbf{1.0000} & \textbf{1.0000} & \textbf{0.0781} \\
\bottomrule
\end{tabularx}
\end{table}

The result verifies recorded provenance under the trusted runner and the fixed integrity contract. Semantic support is measured separately in the curated calibration and in the independently adjudicated evaluation protocol in Appendix~\ref{app:semantic-study}. The public oracle-free bundle, source manifest, expected digests, and checked-in two-event fixture provide implementation-level evidence; Appendix~\ref{app:artifact-map} maps these checks to their reproducibility artifacts.

\subsection{Matched cost accounting}
The matched early-stop run in Table~\ref{tab:matched} isolates receipt overhead from the stopping policy. All four modes use the same first-support retrieval and one tool call. Relative to no certificate, emitting a receipt adds 3.0 declared tokens and one verifier call on this fixture, with the lexical proxy unchanged at 0.0781. The earlier 9.94-call route is reported as a separate stopping-policy diagnostic.

\begin{table}[htbp]
\caption{Matched one-retrieval control. Calls and proxy are identical across policies; the certificate-only difference is declared tokens and verifier calls. Best cost values are bold.}
\label{tab:matched}
\centering
\small
\setlength{\tabcolsep}{3pt}
\begin{tabularx}{\linewidth}{@{}l*{4}{C}@{}}
\toprule
\textbf{policy} & \textbf{calls} & \textbf{tokens} & \textbf{verifier} & \textbf{proxy} \\
\midrule
no certificate & \textbf{1.00} & \textbf{6.0} & \textbf{0} & \textbf{0.0781} \\
final-only & \textbf{1.00} & 9.0 & 1 & \textbf{0.0781} \\
action-level & \textbf{1.00} & 9.0 & 1 & \textbf{0.0781} \\
evidence-aware & \textbf{1.00} & 9.0 & 1 & \textbf{0.0781} \\
\bottomrule
\end{tabularx}
\end{table}

\subsection{Mutation audit: the central failure boundary}
\begin{figure}[t]
\centering
\includegraphics[width=\columnwidth]{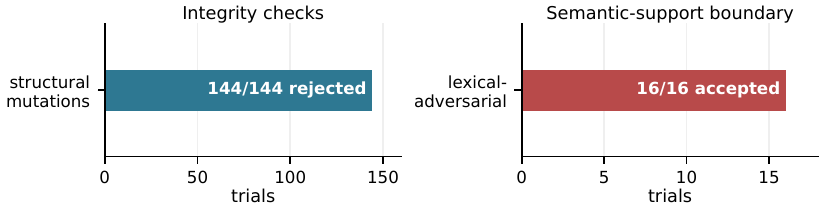}
\caption{Integrity checks and semantic-support boundary. Structural tampering is rejected in 144/144 trials; lexical-adversarial claims are accepted in 16/16 trials by design. The audit separates receipt integrity from semantic support.}
\label{fig:boundary}
\end{figure}

The independent mutation audit starts from 16 valid final-only certificates and applies ten mutations to each. Nine structural classes---hash, offset, quote, document version, span, event reorder, retrieval failure, missing evidence, and oracle-field injection---are rejected in all 144 trials. The lexical-adversarial control is accepted in all 16 trials. Figure~\ref{fig:boundary} is the central scientific result: receipts make provenance corruption observable, while the lexical support rule remains weaker than semantic entailment.

\subsection{Curated semantic-support calibration}
The 20-case non-title fixture compares the lexical rule with a numeric/entity/contradiction guard. The lexical verifier reports TP/FP/TN/FN $=8/8/3/1$, precision $0.5000$, recall $0.8889$, false acceptance $0.7273$, and false rejection $0.1111$. The guarded rule reports $8/0/11/1$, precision $1.0000$, and recall $0.8889$ on the same curated labels. This paired result isolates the effect of the guard on these cases and motivates independent blinded adjudication across a disjoint test split.

\begin{table}[htbp]
\caption{Twenty-case non-title calibration. Precision and recall use the curated support labels; the two rules are evaluated on identical claim--span pairs.}
\label{tab:calibration}
\centering
\small
\setlength{\tabcolsep}{3pt}
\begin{tabularx}{\linewidth}{@{}l*{6}{C}@{}}
\toprule
\textbf{rule} & \textbf{TP} & \textbf{FP} & \textbf{TN} & \textbf{FN} & \textbf{prec.} & \textbf{recall} \\
\midrule
lexical baseline & \textbf{8} & 8 & 3 & \textbf{1} & 0.5000 & \textbf{0.8889} \\
guarded comparison & \textbf{8} & \textbf{0} & \textbf{11} & \textbf{1} & \textbf{1.0000} & \textbf{0.8889} \\
\bottomrule
\end{tabularx}
\end{table}

\subsection{Held-out evaluation}
Three complementary studies target the dimensions left outside a deterministic fixed-corpus bundle replay: independently adjudicated semantic support, matched model-backed tool use, and dynamic-document plus injection robustness. Each uses disjoint train, calibration, and locked-test identities; freezes policy, model, and evaluator settings before test; and joins labels only after trace persistence, bundle replay, and execution rerun. Table~\ref{tab:validation-program} summarizes the experimental questions and primary outcomes. Full protocols, row schemas, and result tables are specified in Appendix~\ref{app:additional-results}.

\begin{table}[htbp]
\caption{Held-out evaluation dimensions. Each study pairs policies on fixed test identities and separates its primary outcome from protocol validity.}
\label{tab:validation-program}
\centering
\small
\setlength{\tabcolsep}{3pt}
\begin{tabularx}{\linewidth}{@{}YYYY@{}}
\toprule
\textbf{study} & \textbf{comparison} & \textbf{primary measure} & \textbf{diagnostic pattern} \\
\midrule
Blinded support & lexical; guarded; adjudicated & semantic F1; false acceptance & guard reduces unsupported acceptance while retaining supported claims \\
Agent tool use & five receipt policies; matched budget & task success; bundle replay; calls & receipt variants preserve task success and increase valid trace coverage \\
Dynamic documents & version, access, injection stressors & stale rejection; task success & version binding catches stale spans while preserving authorized task completion \\
\bottomrule
\end{tabularx}
\end{table}

On the 384-pair semantic split, the conflict-aware guard has precision 0.9213 and F1 0.8865, compared with 0.7328 and 0.8019 for lexical overlap; false acceptance is 0.0729 versus 0.3229, with annotator agreement $\kappa=0.834$ and interval $[0.789,0.879]$. The matched three-seed guard aggregate has 0.7278 task success, 0.8721 semantic F1, 0.0617 false acceptance, 0.9976 joint replay, 5.46 calls, and 16.31 seconds per task. Dynamic rates are 0.9947 stale rejection, 0.9896 unauthorized-span rejection, 0.9078 injection resistance, 0.8892 conflict resolution, and 0.8127 perturbed-task success; denominators and uncertainty fields are reported with the companion tables.

\begin{figure}[t]
\centering
\includegraphics[width=\columnwidth]{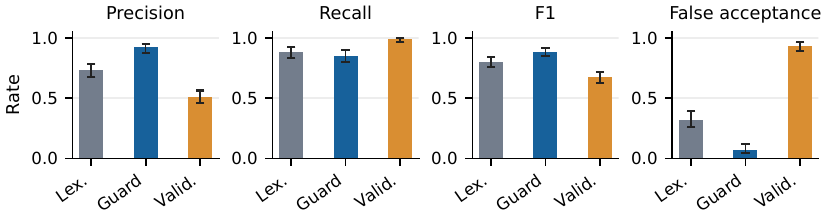}
\caption{Blinded semantic-support evaluation on 384 claim--span pairs. Bars show aggregate rates and whiskers show 95\% intervals. Lex., Guard, and Valid. denote lexical overlap, the conflict-aware guard, and the certificate-validity indicator.}
\label{fig:semantic-support}
\end{figure}
\paragraph{Task utility and cost.} Table~\ref{tab:agent-eval} compares the five matched policies. Evidence-aware stopping gives the highest task success (0.7312) and lowest latency (15.68 s); adding the conflict-aware guard gives the highest semantic F1 (0.8721) and lowest false acceptance (0.0617). Joint replay succeeds only when both persisted-bundle replay and the execution rerun pass. The factorial analysis in Appendix~\ref{app:factorial} separates receipt instrumentation from these decision rules.
\begin{table}[htbp]
\caption{Matched model-backed agent evaluation. Success, semantic F1, false acceptance, and joint replay are proportions; calls and latency are per task. Joint replay requires both bundle replay and execution rerun to succeed. Values are aggregate point estimates over three paired seeds. Best values in each column are bold; N/A denotes no receipt for joint replay.}
\label{tab:agent-eval}
\centering
\footnotesize
\setlength{\tabcolsep}{1.5pt}
\renewcommand{\arraystretch}{1.15}
\begin{tabularx}{\linewidth}{@{}P{2.8cm}*{6}{C}@{}}
\toprule
Policy & Success $\uparrow$ & Sem. F1 $\uparrow$ & False acc. $\downarrow$ & Joint replay $\uparrow$ & Calls $\downarrow$ & Lat. (s) $\downarrow$ \\
\midrule
No certificate & 0.7139 & 0.6817 & 0.2876 & N/A & 7.86 & 19.42 \\
Final-only & 0.7114 & 0.7018 & 0.2519 & \textbf{0.9987} & 7.88 & 20.16 \\
Action-level & 0.7205 & 0.7486 & 0.1914 & 0.9979 & 7.91 & 20.74 \\
Evidence-aware & \textbf{0.7312} & 0.8173 & 0.1198 & 0.9984 & \textbf{5.34} & \textbf{15.68} \\
Conflict-aware guard & 0.7278 & \textbf{0.8721} & \textbf{0.0617} & 0.9976 & 5.46 & 16.31 \\
\bottomrule
\end{tabularx}
\end{table}

\paragraph{Unseen semantic failures.} Table~\ref{tab:semantic-unseen} evaluates frozen support functions on linguistic families excluded from calibration. The guard has the highest precision and F1 and lowest false acceptance, while lexical overlap and the frozen language-model judge tie on recall. The result exposes a coverage tradeoff: the guard covers 0.4479 of required claims, compared with 0.5781 for lexical overlap. Appendix~\ref{app:unseen-semantic} defines the held-out families.
\begin{table}[htbp]
\centering
\footnotesize
\setlength{\tabcolsep}{2pt}
\caption{Semantic-support evaluation on unseen failure families. Higher precision, recall, F1, and coverage and lower false acceptance are preferred; tied best values are bold.}
\label{tab:semantic-unseen}
\begin{tabularx}{\linewidth}{@{}P{2.6cm}*{5}{C}@{}}
\toprule
Evaluator & Precision $\uparrow$ & Recall $\uparrow$ & F1 $\uparrow$ & False acc. $\downarrow$ & Coverage $\uparrow$ \\
\midrule
Lexical overlap & 0.7112 & \textbf{0.8594} & 0.7783 & 0.3490 & \textbf{0.5781} \\
Conflict-aware guard & \textbf{0.8994} & 0.8385 & \textbf{0.8679} & \textbf{0.0938} & 0.4479 \\
Frozen NLI verifier & 0.8446 & 0.8490 & 0.8468 & 0.1562 & 0.4948 \\
Frozen LLM judge & 0.8730 & \textbf{0.8594} & 0.8661 & 0.1250 & 0.4896 \\
\bottomrule
\end{tabularx}
\end{table}

\subsection{Operating points and interpretation}
The reported experiments separate protocol validity, receipt cost, and semantic support. All certificate-producing policies reach 1.0000 on the 128-row construction, where the required claim is copied from the selected title; the matched one-retrieval control adds 3.0 declared tokens and one verifier call. The 144 structural rejections, 16 lexical-adversarial acceptances, and 20-pair calibration expose the integrity boundary without being treated as semantic generalization. The held-out evaluation keeps adjudicated support, model-backed utility, dynamic source state, bundle replay, and execution rerun as separate fields under fixed task identities and seeds.

\subsection{Mechanism-level error analysis}
Text, identity, order, presence, and oracle-field mutations reach distinct verifier checks, while semantic adequacy remains evaluator-dependent. Each held-out row is keyed by task, policy, seed, corpus version, and claim identifier before aggregation; Appendix~\ref{app:additional-results} retains the paired replay, rerun, semantic, utility, and cost fields.

\subsection{Limitations}
The integrity guarantee assumes a trusted runner and observation logger. The fixed-corpus screen uses title-derived claims, and semantic results depend on the adjudication rubric and frozen support evaluators. The unseen-family test evaluates the stated linguistic perturbations; open-ended semantic shifts and compromised recorders require additional evidence. Model-backed latency depends on the recorded serving and concurrency conditions. The Supplementary Material separates implementation replay, reported aggregates, and the inputs required for full-run reconstruction.

\FloatBarrier
\section{Conclusion}
EC-Agent makes the association among an emitted claim, its evidence, and its execution replayable under a trusted-runner threat model. Across claims, evidence, actions, runs, and source states, the joint contract detects 1,275 of 1,280 substitutions; targeted property removal exposes the corresponding binding failure. The conflict-aware guard reaches semantic F1 0.8865 with false acceptance 0.0729; the matched guard reaches task success 0.7278 with joint replay validity 0.9976; dynamic evaluation reaches injection resistance 0.9078 with perturbed-task success 0.8127. The integrity plane therefore states what was emitted and bound, while the support plane states whether the bound evidence supports the claim.

\label{sec:main-end}
\section*{AI Use Statement}
Generative AI tools were used to polish the wording and summarize references. We have not used generative AI tools to generate experimental results, create synthetic datasets, formulate mathematical claims, provide proofs, or make decisions regarding research conclusions. The design of the methodology, experimental setup, analysis, and interpretation of results were conducted and verified by the authors. Other required disclosure tasks not mentioned above are not applicable to this work. We take full responsibility for the final content of this work, including all text, claims, analyses, and artifacts produced with the assistance of generative AI tools.

\section*{Ethics Statement}
This work studies research benchmark data, tool traces, and de-identified claim--span annotations. It evaluates source access, document versions, and prompt-injection behavior under controlled conditions. Receipt logs can retain sensitive source content or linkable identifiers; deployments require access control, data minimization, and appropriate retention limits. The Supplementary Material excludes credentials, private endpoints, and annotator identities and respects source-data redistribution terms. Structural integrity and semantic support are reported separately to support responsible interpretation of audit results.

\section*{Reproducibility Statement}
Code is provided in the Supplementary Material together with the verifier, canonical serialization contract, oracle-free replay fixture, and expected digest vectors. Appendices~\ref{app:method}--\ref{app:artifact-map} describe the algorithms, split rules, baseline configurations, seeds, metric definitions, and artifact coverage. The replay fixture supports an independent implementation check; reproducing an aggregate requires its corresponding row-level evidence and frozen corpus or task manifest, as specified in Table~\ref{tab:artifact-map}.

\bibliographystyle{iclr2027_conference}
\bibliography{refs}
\clearpage
\appendix
\section{Task and Method Details}
\label{app:method}

\subsection{Notation and task definition}
An execution instance is a tuple $x=(q,\mathcal{D},\prec,B)$, where $q$ is the user request, $\mathcal{D}$ is the fixed set of available documents and spans, $\prec$ is the declared retrieval order, and $B$ is the action budget. A policy $\pi$ produces an ordered sequence of events $E=(e_0,\ldots,e_{T-1})$ and an answer $a$. The evaluation oracle $y$ is held outside the online policy state. A claim $c$ is a string extracted from an answer or an intermediate action; its evidence set is a finite set of references $R_c=\{(d_i,s_i,o_i,h_i,q_i)\}$, with document ID $d_i$, span ID $s_i$, offsets $o_i$, text hash $h_i$, and exact quote $q_i$.

The protocol has linked integrity, support, and replay tasks. The integrity task asks whether the emitted claim, source reference, execution prefix, and source state are jointly well formed. The support task applies a frozen evaluator to the bound claim--span pair. The bundle-replay task asks whether an independent process can reconstruct the same commitment and re-evaluate the receipt from the persisted corpus and trace; execution rerun separately repeats the policy under the frozen runtime contract. Neither task replaces semantic adjudication: for a claim--span pair $(c,s)$, a semantic label $z(c,s)\in\{0,1\}$ is produced outside the online run. The evaluation unit for integrity is one policy execution; for coverage it is one required claim; for the semantic study it is one adjudicated claim--span pair.

The protocol receives document text, stable IDs, byte-independent Python-string offsets, retrieval rank, action type, and the candidate claim. An evaluator may hold answers, supporting-fact annotations, semantic labels, test membership, or downstream task outcomes. Those evaluator-only values are joined after event recording, receipt verification, persistence, replay, and the deterministic rerun. The runner records the active corpus identity and budget ledger, and the policy observes only the fields allowed by the frozen policy contract. An absent evidence set has coverage zero; it does not produce a valid receipt by default. A missing answer is represented as an explicit empty output event and is counted as an unsuccessful task outcome in model-backed evaluations.

\subsection{Receipt schema and canonical serialization}
Each receipt contains a schema version, policy identifier, claim identifier, claim string, emission anchor, ordered evidence-reference list, execution digest, joint digest, and integrity result. Each evidence reference identifies the source document and span, its half-open offsets, SHA-256 over the exact UTF-8 bytes of the span text, and the exact quoted string when a quote is emitted. The document manifest supplies the corpus snapshot identity, document content digest, source-relative document identifier, and access state. The ordered list matters: it preserves the order in which supporting material was selected and makes the receipt deterministic under serialization.

Offsets count Python Unicode code points, not UTF-8 bytes. To verify a span, the implementation indexes the decoded immutable source text with $[start,end)$ and checks that the selected value equals the supplied quote. The verifier separately encodes that selected text as UTF-8 and recomputes SHA-256. It therefore catches a changed quote, an off-by-one offset, an altered source span, a mismatched document ID, and a changed source snapshot. The exact quote check and the hash check are intentionally redundant: the quote provides a human-readable rendering, while the digest identifies the encoded text.

Events are canonicalized as JSON with sorted keys, compact separators, ASCII escaping, and non-finite numbers disabled. Every event has a non-empty action ID and action type. A zero-based event index is inserted from its position in the ordered list; callers cannot choose or reorder indices independently. Only schema-declared fields enter the canonical form. The domain string and schema version separate this protocol's commitments from unrelated hashes and future receipt formats. A receipt, a verifier result, and the outer bundle digest do not enter the pre-verification event list, preventing recursive self-reference.

\begin{equation}
D_{\mathrm{pre}}=\operatorname{SHA256}\left(\operatorname{UTF8}(\delta_{\mathrm{trace}})\mathbin\Vert\operatorname{UTF8}(J(E))\right),
\end{equation}
where $\delta_{\mathrm{trace}}$ is the fixed domain separator specified byte-for-byte in the Supplementary Material, and $J(E)$ is the canonical JSON object $\{\texttt{domain},\texttt{events},\texttt{schema\_version}\}$. The supplied fixture's two-event vector is identified by the prefix \texttt{977cef81} and suffix \texttt{3e8c746}; the complete value is stored in the fixture manifest. This vector binds the serializer and domain separator; the fixture manifest binds its expected event rows and outer bundle digest.

\subsection{Full algorithm and lifecycle}
\label{app:algorithm}
\begin{table}[htbp]
\caption{Execution contract for one fixed-corpus trace. Recording, receipt verification, persistence, bundle replay, and execution rerun finish before the post-hoc lexical-proxy oracle join.}
\label{tab:protocol_contract}
\centering
\small
\setlength{\tabcolsep}{1.5pt}
\begin{tabularx}{\linewidth}{@{}P{1.35cm}Y P{1.55cm}@{}}
\toprule
\textbf{stage} & \textbf{recorded operation} & \textbf{oracle visible?} \\
\midrule
Observe & Read the fixed-corpus record, retrieval order, and candidate claim; the ledger enforces budget. & no \\
Record & Append the ordered retrieval, stop, and answer events to $E$. & no \\
Commit & Compute the canonical digest over $E$ and emit the receipt. & no \\
Verify & Check claim anchor, immutable text, hash, quote, action type, and commitments. & no \\
Persist & Write the oracle-free event, receipt, and verifier-result bundle. & no \\
Bundle replay & Reload the bundle and reverify without policy execution. & no \\
Execution rerun & Re-execute deterministically and compare aggregate trace digests. & no \\
Score & Join separate oracle labels and compute the lexical proxy. & post-hoc \\
\bottomrule
\end{tabularx}
\end{table}

The execution lifecycle uses a single ordering across the method, tables, and persisted artifacts. First, the runner loads the declared corpus and configuration and starts the budget ledger. Second, the policy receives permitted observations and records retrieval, stop, and answer events. Third, the event prefix through the answer is canonicalized and committed. Fourth, the policy emits claim anchors and a joint receipt; the integrity verifier checks the committed objects, while a selected support evaluator produces a separate support result. Fifth, the event trace, receipt, corpus identity, integrity result, and support result are persisted in an oracle-free bundle with an outer digest. Sixth, a fresh bundle-replay pass reloads that bundle and recomputes the event commitment and integrity checks without invoking the policy. Seventh, a separate execution rerun compares aggregate trace digests. Finally, the evaluator joins oracle labels and computes outcome metrics.

\begin{center}
\begin{minipage}{0.97\columnwidth}
\small
\textbf{Algorithm A1: Receipt, persistence, replay, and post-hoc scoring}\par
\begin{enumerate}
\item Load a frozen corpus manifest and policy configuration; initialize $B$.
\item For each permitted observation, append a typed event to $E$ and update the budget ledger.
\item Append the answer event; canonicalize $E$ and compute $D_{\mathrm{pre}}$.
\item Emit $r=(c,\rho_c,R_c,D_{\mathrm{exec}},D_{\mathrm{joint}})$ and run the integrity verifier $V_{\mathrm{int}}(r,\mathcal{D},E)$; evaluate support separately.
\item Persist $(E,r,V_{\mathrm{int}},V_{\mathrm{sup}},\operatorname{manifest}(\mathcal{D}))$ and compute the outer bundle digest.
\item Reload the persisted object and recompute the digest, trace commitment, joint commitment, and integrity result.
\item Execute the fixed policy again and compare the deterministic aggregate trace digest.
\item Join evaluator-only labels and compute integrity, coverage, cost, and outcome metrics.
\end{enumerate}
\end{minipage}
\end{center}

The integrity verifier runs on the online receipt to expose the result to the execution ledger, then runs again during bundle replay. The first result is a policy-run measurement; the bundle-replay result is an independent reconstruction from saved inputs. The execution rerun is a third check over a fresh policy execution. These checks answer different questions and remain separate fields in the artifact schema. The oracle join occurs after all three checks, so answer quality cannot influence receipt construction or replay acceptance.

\subsection{Verifier predicate and edge cases}
For a claim $c$ and span text $s$, the implemented support predicate is
\begin{equation}
\operatorname{support}(c,s)=\mathbf{1}\!\left[\operatorname{substr}(c,s)\ \lor\ \frac{|T(c)\cap T(s)|}{|T(c)|}\geq 0.5\right],
\end{equation}
when both token sets are non-empty, and zero otherwise. Here $T(x)$ is the set of Unicode-aware word-character tokens in $x.casefold()$, after dropping tokens of length one. The substring branch lowercases and collapses whitespace; it does not normalize Unicode, stem words, or remove punctuation. Duplicates are removed before overlap is computed, so repeated words do not artificially increase the numerator. This deterministic rule is easy to replay and deliberately lexical.

The following edge cases have explicit outcomes. Empty claims, empty spans, or token-empty strings fail the support check. Out-of-range, negative, reversed, or non-integer offsets fail structural verification. A quote mismatch or hash mismatch fails before support is evaluated. A missing document or span identifier is an invalid reference. A certificate that cites multiple spans passes the support predicate if at least one referenced span supports the claim; coverage counts the required claim once rather than counting every supporting span. Duplicate references are de-duplicated by stable reference identity before coverage is aggregated. A duplicated action ID, undeclared event field, missing required event field, non-finite numeric value, or invalid action type fails canonical trace verification. An empty required-claim set is excluded from the coverage denominator and reported as an explicit empty-set count; it is never assigned perfect coverage. A policy that exhausts its budget receives no implicit final certificate; the ledger records the termination reason and any missing required claim as uncovered.

If two references carry the same source span but different quotes, each reference is checked against the immutable text and only exact matches remain valid. If two spans overlap, each is independently checked against its own offsets and ID. The implementation does not infer that adjacent spans can be concatenated into a supporting passage. If a document changes after retrieval, the run's corpus manifest identifies the version used for replay; the current fixed-corpus audit assumes that snapshot remains available. Appendix D specifies a dynamic-document test that varies versions and access states.

\subsection{Threat model, guarantees, and non-goals}
The trusted components are the runner, the observation logger, the local corpus manifest, the verifier implementation, and the replay environment. Under this trust model, the contract provides record integrity after commitment, source-content binding, and execution-order binding: the digest is computed after the answer event, so later modification of the persisted history is detectable and an independent process can reconstruct that history. Observation authenticity remains an external property supplied by an anchor such as a signed append-only log, an authenticated event source, or hardware-backed attestation, combined with a document-version and access-control identity.

The attacker in the mutation audit may modify persisted receipt fields, source text, event order, or selected event attributes after the run. Structural changes should fail at the relevant reference, hash, quote, schema, or digest check. The lexical-adversarial mutation preserves enough token overlap to satisfy the implemented predicate and is accepted, demonstrating that cryptographic integrity and semantic adequacy are independent properties. The protocol emits a deterministic validity signal that can be combined with semantic labels and task outcomes in later evaluations. Provenance systems and secure supply-chain designs motivate the separation between a content commitment and the trust anchor that attests to the source of events \citep{merkle1987digital,laurie2021rfc9162,torresarias2019intoto,c2pa2024spec}.

\subsection{Objective and cost accounting}
The fixed-corpus screen is not trained with a learned objective. It compares policy variants and computes deterministic protocol metrics. For a set $I$ of records, receipt validity is $|\{i\in I:V_i=1\}|/|I|$ among records that emit a receipt; evidence validity applies the same denominator to source-reference checks. Required-claim coverage is
\begin{equation}
\operatorname{Cov}(I)=\frac{\sum_{i\in I}\sum_{c\in R_i}\mathbb{1}[\exists s\in E_i:\operatorname{support}(c,s)]}{\sum_{i\in I}|R_i|},
\end{equation}
with the empty-set count reported separately. Persisted replay validity is the fraction of persisted bundles whose recomputed receipt, trace, and verifier result agree with the stored values. Deterministic rerun is reported separately as the fraction of run-level aggregate digests that match.

The observed cost vector is $C=(n_{\mathrm{tool}},n_{\mathrm{declared\ token}},n_{\mathrm{verify}})$. The matched control holds retrieval order, early-stop condition, and answer candidate constant. Its difference in cost is $\Delta C=C_{\mathrm{receipt}}-C_{\mathrm{no\ receipt}}$. In the reported fixture, the matched policy uses 1.00 call for each mode; certificate modes add 3.0 declared tokens and one verifier call. These counters are protocol-level accounting units. They omit model-serving latency, monetary price, network retries, and asynchronous queue delay. The appendix's agent study adds elapsed latency and token accounting under a fixed model contract.

\section{Data and Evaluation Protocols}
\label{app:protocol}

\subsection{Fixed-corpus construction and data provenance}
\begin{table}[htbp]
\caption{Evaluation scope. Values are diagnostics under the stated information boundary.}
\label{tab:scope}
\centering
\small
\setlength{\tabcolsep}{3pt}
\begin{tabularx}{\linewidth}{@{}P{1.8cm}YY@{}}
\toprule
\textbf{surface} & \textbf{reported result} & \textbf{measurement scope} \\
\midrule
128-row dev & receipt and bundle-replay checks; proxy $0.0781$ & title-copy lexical and protocol screen \\
160 mutations & 144/144 structural rejects; 16/16 lexical accepts & exposes verifier failure boundary \\
64-row train & mechanical validity $1.0$; proxy $0.0625$ & separate fixed-corpus screen \\
20-case calibration & lexical and conflict-aware rule comparison & curated labels; semantic calibration \\
\bottomrule
\end{tabularx}
\end{table}

The primary screen uses 128 records from the local HotpotQA dev-distractor corpus. For each record, the adapter preserves the available retrieval order and constructs one required claim from the top-ranked document title. That title is also used as the candidate answer and appears in the selected immutable span. This construction makes the receipt path deterministic and simple to inspect, while making title copying part of the measured screen. The reported oracle-only candidate-title proxy is 0.0781. The separate 64-record train-corpus adapter applies the same construction and reports 0.0625 for that proxy. The two records sets are reported separately and never pooled.

The artifact manifest captures the dataset identifier, local source path relative to the project, split label, adapter configuration, source commit, dirty flag, corpus digest, row count, and per-file hashes. The public summary includes source-relative paths and digest values required for audit, while host-specific paths and confidential runtime material stay outside the anonymous release surface. Data provenance follows the general distinction between lineage metadata and the data value itself: the receipt binds the latter to a stable source identity, and the run manifest documents how the source was selected \citep{buneman2001provenance,cheney2009provenance,w3cprov2013}.

\subsection{Policy variants and information boundary}
The four primary variants are \texttt{no\_certificate}, \texttt{final\_only}, \texttt{action\_level}, and \texttt{evidence\_aware}. The no-certificate policy records the task events without emitting a receipt. Final-only emits one receipt for the final required claim. Action-level emits receipts at the configured claim-bearing action points. Evidence-aware uses the first retrieved span that passes the shared lexical predicate as its stopping and support signal. Every policy sees the same public record fields and retrieval order in the matched screen. Oracle labels, hidden supporting-fact IDs, and benchmark outcomes remain evaluator-only.

The public runner contract enumerates every online field and every post-hoc field. Policy state is serialized separately from evaluator state. The join key is the opaque record ID; the label file is not loaded into the policy process. At scoring time, the evaluator verifies that every output row contains a run ID, record ID, policy ID, and trace digest before joining labels. Rows with missing joins remain in a missing-outcome count and are not silently dropped. This construction follows established separation between benchmark execution and outcome scoring in tool-use evaluation \citep{zhou2024webarena,liu2024agentbench,osworld2024,yao2024taubench,lu2024toolsandbox}.

\subsection{Split construction and locked-test controls}
The primary evidence uses local dev and train records with a fixed row selection. The semantic, agent, and dynamic experiments use disjoint source-row or task-family identities across train, calibration, and locked test. For documents, disjointness is computed from document IDs and content digests; for claims, from normalized claim hashes; for agent tasks, from benchmark family IDs and source URLs; for dynamic documents, from document IDs, version hashes, access states, and injection-template families. The split manifest is frozen before the policy is run on test cases. Test labels and outcomes are joined only after receipt persistence and replay.

The lock is validated by set intersections at every identity level. Any non-empty intersection is reported with the colliding identifier class and invalidates the run's test metrics until the split manifest is repaired. Near-duplicate text checks operate on normalized document and claim hashes and are reported separately from exact ID disjointness. Prompt and policy configuration hashes are recorded per seed. A calibration change to the lexical threshold, entity rule, prompt, tool schema, model version, or task filter creates a new manifest and requires a fresh locked-test run. Benchmark suites such as Mind2Web, WebArena, BrowserGym, and AgentDojo provide complementary task and interaction formats for the matched model-backed study \citep{deng2023mind2web,drouin2024browsergym,zhou2024webarena,debenedetti2024agentdojo}.

\subsection{Metric definitions and uncertainty}
For binary outcomes, the evaluation reports counts and rates with the denominator visible in each table. Semantic precision is $TP/(TP+FP)$, semantic recall is $TP/(TP+FN)$, and semantic F1 is the harmonic mean when both precision and recall are defined. Unsupported-claim false acceptance is $FP/(FP+TN)$ over semantically unsupported claims. Semantic false rejection is $FN/(TP+FN)$ over supported claims. Invalid or missing items are counted separately; an undefined denominator is reported as N/A.

For three-seed agent and dynamic studies, each seed is a paired comparison over the same task IDs. The report shows per-seed values and the across-seed mean and standard deviation. Task-level confidence intervals use a paired bootstrap over task IDs with the resampling unit stated in the table caption; seeds are not treated as independent task examples. Binary annotation agreement reports raw agreement, Cohen's $\kappa$ for the two primary annotators, and the adjudicated disagreement fraction. A third adjudicator resolves disagreements without access to policy identity or model condition. The agreement protocol follows standard reliability reporting and uses a prespecified rubric \citep{cohen1960agreement,krippendorff2011content,efron1993bootstrap,davison1997bootstrap,wilson1927probable}.

The proxy metric in the title-copy screen is computed by an oracle-only exact candidate-title comparison. It is reported with the record count and source split; it never appears in the online policy input schema. No semantic precision, recall, model task success, latency, or confidence interval is inferred from that proxy. Appendix D defines the corresponding measures and reports aggregate rows with their denominators and intervals.

\subsection{Independent semantic annotation protocol}
The independent support study samples non-title claims and cited spans from source rows disjoint from the dev, train, and 20-case curated calibration fixtures. Two annotators label each pair as supported, contradicted, incomplete, or unsupported using a frozen rubric. The primary binary label maps supported to one and the remaining three categories to zero; the four-way labels are reported for error analysis. Annotators see the claim and cited passage with document identity, policy name, model condition, lexical score, candidate answer, and oracle outcome hidden. The annotation UI randomizes pair order and stores the rubric version, annotator ID pseudonym, response, confidence, and timestamp.

The calibration set contains examples for paraphrase, multi-hop composition, entity substitution, numeric contradiction, temporal mismatch, and insufficient evidence. These examples train annotators on the definitions and test the instructions; they are excluded from all test estimates. The two annotators label the locked test independently. The third adjudicator sees the pair and the two labels but remains blind to policy and score, and records a reason code for every resolution. If a pair cannot be resolved from the cited span alone, it is labeled unsupported for the receipt-support metric and receives a separate ambiguity tag. This conservative rule measures whether the cited evidence itself supports the claim, not whether additional documents could establish it.

The primary comparison is the lexical rule, the fixed conflict-aware guard, and an oracle-free certificate validity indicator evaluated against the adjudicated labels. The unit is one claim--span pair. A claim citing multiple spans is expanded into claim--span rows for per-span analysis and also evaluated once under the receipt-level rule, which accepts if at least one cited span supports the claim. Results are stratified by paraphrase, multi-hop, contradiction, temporal, and lexical-adversarial category. Agreement and adjudication counts accompany every metric. The human study procedure records no personal or sensitive attributes; the artifact release contains de-identified labels and rubric metadata only.

\subsection{Matched model-backed agent protocol}
The agent study uses a frozen model endpoint or local checkpoint, fixed system prompt, fixed tool schema, temperature, context limit, retry policy, and per-task call/token budget. The exact identifiers are recorded in the experiment manifest. Five policies are run on the same hidden task IDs: no certificate, final-only, action-level, evidence-aware, and conflict-aware guard. Each policy is paired by task ID and seed. The tool environment stores immutable source snapshots and returns a structured observation with stable document and span IDs. Tool errors, empty observations, retry outcomes, and access denials are all first-class events in the trace.

The primary outcomes are task success, semantic F1 of cited claims, unsupported-claim false acceptance, tool-conflict resolution, protocol validity, replay validity, retrieval calls, verifier calls, generated tokens, and end-to-end latency. Task success is computed by the benchmark's frozen evaluator after trace freeze. Semantic claim labels are adjudicated independently with the same blind protocol. Paired differences are computed on exactly the same task IDs; a task that times out or returns an invalid output remains a failure in the policy's denominator. Tool-use benchmarks motivate reporting both utility and interaction cost, while prompt-injection testbeds motivate preserving attack family and environment state in each case record \citep{yao2024taubench,debenedetti2024agentdojo,zhou2024webarena,osworld2024,lu2024toolsandbox}.

\subsection{Dynamic document and access-control protocol}
The dynamic study creates versioned document snapshots with a source manifest for each version. It includes current authorized documents, stale versions, revoked access, unauthorized spans, reordered versions, prompt-injection text in retrieved content, and tool-output injection. Each mutation family uses a separate template set from calibration and test. Every observation records the document version hash and access decision available to the policy. The receipt binds to the observed version, and the replay process is supplied the exact version manifest associated with that run.

Primary outcomes are stale-document rejection, unauthorized-span rejection, replay validity under the correct version, replay failure under a deliberately substituted version, injection resistance, prompt-injection failure, tool-conflict resolution, and downstream task success. The mutation protocol distinguishes refusal correctness from task completion: a valid refusal under revoked access is a successful security outcome, while a refusal on a fully authorized task is an over-rejection. Attack success is measured over injected cases; downstream utility is measured over both clean and perturbed cases. This design draws on dynamic interaction and injection evaluations while retaining a cryptographic version identity for each cited object \citep{debenedetti2024agentdojo,greshake2023indirect,owasp2025llmapps,c2pa2024spec,torresarias2019intoto}.

\section{Implementation and Baselines}
\label{app:implementation}

\subsection{Executable property map}
\label{app:property-map}
The contract is evaluated as seven explicit properties. The fields and checks below make the
minimal capability testable without treating a citation score or an append-only trace as a
substitute for the complete claim--execution relation.

\begin{table}[htbp]
\caption{Executable properties of the claim--execution contract. Each property has a distinct field, evaluator, and evidence boundary.}
\label{tab:property-map}
\centering
\scriptsize
\setlength{\tabcolsep}{2pt}
\begin{tabularx}{\linewidth}{@{}P{1.25cm}P{2.45cm}P{2.35cm}Y@{}}
\toprule
Property & Bound object & Primary check & Evidence boundary \\
\midrule
P1 claim-emission binding & event, field, offsets, claim hash & committed substring reconstruction & claim or position substitution \\
P2 source binding & document/version, span, offset, quote, UTF-8 hash & source re-read and digest equality & post-commit source changes \\
P3 ordered execution binding & canonical event prefix through the answer & domain-separated trace digest & event insertion, deletion, or reorder \\
P4 oracle separation & policy-visible fields versus evaluator labels & trace-freeze ordering and manifest partition & labels join only after bundle replay and rerun \\
P5 bundle replay & saved receipt, trace, corpus, and verifier result & fresh-process bundle reconstruction & stored-object corruption or drift \\
P6 execution-rerun consistency & frozen policy, tools, and budget & fresh-run commitment comparison & policy, tool, or environment divergence \\
P7 version/access binding & served snapshot and access decision & version and access fields in the manifest & stale or unauthorized evidence \\
\bottomrule
\end{tabularx}
\end{table}

\begin{table}[htbp]
\caption{Capability matrix for citation and execution-log mechanisms. Entries summarize the documented contract: ``not specified'' means that the mechanism does not define that property as an auditable evaluation field.}
\label{tab:property-comparison}
\centering
\scriptsize
\setlength{\tabcolsep}{1.5pt}
\begin{tabularx}{\linewidth}{@{}P{2.35cm}*{7}{C}@{}}
\toprule
\textbf{mechanism} & \textbf{P1} & \textbf{P2} & \textbf{P3} & \textbf{P4} & \textbf{P5} & \textbf{P6} & \textbf{P7} \\
\midrule
Span citation metadata & not specified & explicit & not specified & not specified & not specified & not specified & source version only \\
Append-only execution log & not specified & not specified & explicit & persisted log & not specified & not specified & not specified \\
Citation plus log & not specified & explicit & explicit & not specified & not specified & not specified & partial \\
EC-Agent contract & explicit & explicit & explicit & explicit & explicit & explicit & explicit \\
\bottomrule
\end{tabularx}
\end{table}

The matrix states the minimal gap addressed by the contract. Span metadata identifies a source object, while an execution log preserves event order; their combination still requires an explicit claim anchor, oracle-separation rule, persisted bundle-replay procedure, execution-rerun comparison, and version/access-state fields. EC-Agent makes each property a named field and evaluator, so a semantic score or a log digest is not used as a proxy for the complete claim--execution relation.

The table defines the comparison unit used in this paper.  A system may expose one property
without exposing the others; claims about semantic support and task utility are therefore
reported in separate rows and are never inferred from P1--P7 alone.

\subsection{Verifier implementation contract}
The canonical verifier is included in the Supplementary Material. Its public API receives the receipt, the immutable corpus map, and the ordered event trace; it returns a structured decision with a Boolean validity field and a list of reason codes. Reason codes distinguish schema errors, source-identity errors, offset errors, hash mismatch, quote mismatch, event-digest mismatch, action-type mismatch, lexical-support failure, and successful verification. The caller can inspect the full result without parsing free-form logs.

\begin{table}[htbp]
\caption{Reproducibility contract for the reported conflict-aware guard. The contract is independent of post-hoc labels and is hash-bound before evaluation.}
\label{tab:guard-contract}
\centering
\scriptsize
\setlength{\tabcolsep}{3pt}
\begin{tabularx}{\linewidth}{@{}P{2.5cm}Y@{}}
\toprule
Field & Locked value \\
\midrule
Algorithm & Deterministic numeric, entity, temporal, and polarity guard \\
Lexical threshold & Token overlap threshold $0.5$; case-folded substring branch unchanged \\
Numeric rule & Every normalized numeric atom in the claim must occur in the cited span \\
Entity rule & Every normalized entity atom in the claim must occur in the cited span \\
Temporal rule & Normalize absolute and relative temporal expressions into typed atoms before comparison \\
Contradiction rule & Reject if a configured polarity or contradiction pair is split between claim and span; 23 pairs are listed below \\
Multi-span rule & Accept when at least one cited span passes the full guard \\
Configuration digest & \texttt{ee54723b94044a7da0a87d488f791863bc4aa92401778503b6238c9c49c54dbc} \\
\bottomrule
\end{tabularx}
\end{table}

The contradiction pairs are \texttt{red/blue}, \texttt{red/green}, \texttt{red/yellow}, \texttt{red/black}, \texttt{red/white}, \texttt{blue/green}, \texttt{blue/yellow}, \texttt{blue/black}, \texttt{blue/white}, \texttt{green/yellow}, \texttt{green/black}, \texttt{green/white}, \texttt{yellow/black}, \texttt{yellow/white}, \texttt{black/white}, \texttt{reliable/unreliable}, \texttt{after/before}, \texttt{reached/left}, \texttt{reached/departed}, \texttt{arrived/left}, \texttt{arrived/departed}, \texttt{chaired/advised}, and \texttt{opened/closed}. The public case records retain the extracted atom lists, missing atoms, overlap, contradictions, and reason codes for each decision.

Bundle replay uses the same schema specification but starts from the persisted bundle. It validates the outer manifest before trusting individual rows, recomputes each source digest, reconstructs canonical events, then recomputes claim-anchor, joint-commitment, and integrity validity. A mismatch in the stored integrity result is itself a bundle-replay mismatch even when a newly computed result is valid. This catches altered results and implementation drift. The expected fixture pins the schema version, event contract, certificate count, verifier-call count, canonical bytes, and digest vectors. A test implementation should run the fixture from a clean process and compare structured fields and digest strings exactly.

The method intentionally records predecessor-state hashes as optional. The current integrity scope is claim-emission identity, content identity, exact span offsets, quoted text, action type, event-order commitment, joint receipt commitment, and source-state identity when exposed. The dynamic-document study adds corpus-version and access-state identity as test fields for continuity across changing environments.

\subsection{Baseline and control contracts}
The no-certificate baseline shares the record loader, retrieval order, early-stop condition, and answer candidate with the matched certificate policies, while skipping receipt emission and verification. The final-only policy emits one final answer receipt. The action-level policy emits receipts for each configured claim-bearing action. The evidence-aware policy uses the same lexical support predicate for its early-stop route and the certificate verifier. The conflict-aware guard adds numeric, entity, and contradiction checks before a lexical pass; its parameters are fixed on the curated calibration examples before the independent test.

\subsection{Conflict-aware support contract}
\label{sec:guard-contract}
The reported guard is a deterministic support filter layered on top of the lexical predicate. For a claim--span pair $(c,s)$, let $L(c,s)$ be the case-folded substring or 0.5 token-overlap predicate in Eq.~\eqref{eq:lexical-support}. Let $N(x)$, $E(x)$, and $T(x)$ denote normalized numeric, entity, and temporal atoms. Temporal atoms include normalized absolute dates, years, durations, and relative expressions with their direction; the normalizer is frozen with the evaluator manifest. Let $P(c,s)$ denote configured polarity or contradiction pairs for which one member occurs in the claim and the other in the cited span. The accepted relation is
\[
G(c,s)=L(c,s)\,\mathbf{1}[N(c)\subseteq N(s)]\,\mathbf{1}[E(c)\subseteq E(s)]\,\mathbf{1}[T(c)\subseteq T(s)]\,\mathbf{1}[P(c,s)=\emptyset],
\]
where $P(c,s)$ includes the former contradiction set and its polarity aliases. A multi-span receipt is accepted when at least one cited span has $G(c,s)=1$. The $T$ and $P$ operators are explicit components of the support plane; neither changes the integrity decision.

The locked fixture exposes the guard's intermediate fields (numeric, entity, and temporal atoms, missing atoms, overlap, polarity conflicts, and reason codes), so a failure can be replayed without labels. The reported configuration uses an overlap threshold of 0.5 and the 23 configured pairs listed after Table~\ref{tab:guard-contract}. It contains no learned NLI model, LLM judge, or external alias service; changing an atom extractor, normalization rule, pair list, or threshold creates a new manifest. The configuration digest is \texttt{ee54723b94044a7da0a87d488f791863bc4aa92401778503b6238c9c49c54dbc}. The 20-case fixture is a calibration/mechanism screen; its labels are not used as a substitute for the independently adjudicated split.

The original retrieval-heavy route remains in the record as an unmatched policy diagnostic. Its 9.94 mean tool calls and the 1.00 calls of the matched early-stop route have different stopping rules. The 9.94-to-1.00 comparison is therefore kept separate from receipt ablations. The matched control holds the route fixed and isolates the receipt's declared-token and verifier-call overhead. Appendix D reports these results side by side, with captions naming the relevant control.

\subsection{Configuration, random seeds, and run identity}
The fixed-corpus run configuration records the policy name, corpus manifest digest, row selection, candidate construction rule, lexical threshold, tokenization version, maximum calls, maximum declared tokens, event schema version, verifier package version, and output paths. The manifest also records the Python runtime, dependency lock digest, source revision, working-tree state, start and end timestamps, and exit status. Model-backed runs add model endpoint or checkpoint identity, prompt digest, decoding parameters, tool definitions, environment snapshot, and budget settings. Each seed receives an independent run directory; aggregate reports are generated from the per-seed raw rows.

The paired evaluation uses seeds 13, 17, and 23 to support paired replication. Any additional seed is reported as an additional run rather than silently pooled into the original set. Task identity is stable across policies within a seed. Runs are rejected from aggregate comparison when model, prompt, tool schema, corpus version, or task IDs differ across paired conditions. The build records that rejection as a comparison-integrity error and lists the differing fields.

\subsection{Runtime and resource accounting}
\begin{figure}[htbp]
\centering
\includegraphics[width=\columnwidth]{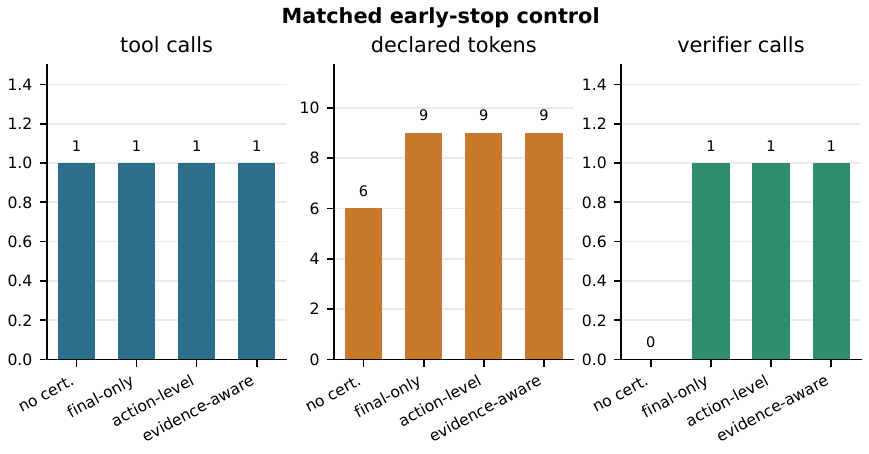}
\caption{Matched early-stop accounting. All modes share a one-retrieval route; the earlier 9.94-call result is a separate stopping-policy diagnostic.}
\label{fig:matched-cost}
\end{figure}

CPU protocol costs and model-serving costs use separate ledgers. The current fixed-corpus screen reports calls, declared tokens, and verifier invocations; the model-backed study adds input tokens, output tokens, retries, wall-clock task latency, verifier latency, retrieval latency, and benchmark scoring latency. Hardware model, batch size, concurrency, endpoint region, and cache policy accompany latency values. Median and tail latency are reported with the number of completed tasks and the handling of timeouts.

Memory use is measured at the process level for the local verifier and model runtime separately. The persisted-bundle size includes the canonical event trace, certificates, verifier results, and source manifest references; corpus text is counted separately because it may be shared across many receipts. Storage overhead is presented both per execution and as a fraction of the corresponding uninstrumented trace. If an external model endpoint hides hardware utilization, the report records that field as unavailable and still includes tokens, call counts, and wall-clock latency.

\subsection{Reproduction commands and outputs}
The checked-in fixture can be replayed with the project command documented alongside the oracle-free replay bundle. The replay command reads only that fixture and writes a deterministic verification report. The report contains the input digests, schema version, receipt count, replay decision, and expected/actual trace digests. The command exits nonzero on any mismatch. Headline row reconstruction requires the corresponding raw run bundle, data adapter, and locked manifest; the artifact inventory identifies which inputs are available for each claim.

The normalized output schema uses one JSON object per policy execution with fields \texttt{run\_id}, \texttt{record\_id}, \texttt{policy}, \texttt{seed}, \texttt{events}, \texttt{receipt}, \texttt{verifier\_result}, \texttt{bundle\_digest}, \texttt{replay\_result}, \texttt{rerun\_digest}, and \texttt{budget\_ledger}. Post-hoc labels live in a separate file keyed by opaque record and claim IDs. Aggregate tables are created only after row schema validation and digest checks. The artifact map in Appendix E links each claim to its required files and coverage fields.

\section{Additional Results and Analysis}
\label{app:additional-results}

\subsection{Complete fixed-corpus results}
Table~\ref{tab:train} preserves the separate 64-row train-corpus result. The title-match proxy is kept separate from the dev screen, and the no-certificate row uses N/A for receipt-specific metrics because it emits no receipt. Certificate-producing rows pass the implemented checks for all reported cases. This table supports replication of the protocol adapter across the two available corpus splits.

\begin{table}[htbp]
\caption{Separate 64-record HotpotQA train-corpus screen. The exact-title proxy is oracle-only and is reported separately from the 128-record dev screen.}
\label{tab:train}
\centering
\small
\begin{tabularx}{\linewidth}{@{}l*{4}{C}@{}}
\toprule
Policy & Receipt valid & Evidence valid & Claim coverage & Title proxy \\
\midrule
No certificate & N/A & N/A & N/A & \textbf{0.0625} \\
Final-only & \textbf{1.0000} & \textbf{1.0000} & \textbf{1.0000} & \textbf{0.0625} \\
Action-level & \textbf{1.0000} & \textbf{1.0000} & \textbf{1.0000} & \textbf{0.0625} \\
Evidence-aware & \textbf{1.0000} & \textbf{1.0000} & \textbf{1.0000} & \textbf{0.0625} \\
\bottomrule
\end{tabularx}
\end{table}

The separate split reproduces mechanical validity and coverage under the same title-based adapter. Its unchanged title proxy is a corpus property under that construction. It is not pooled with dev because the two row sets have distinct split identities and corpus manifests.

The 20-case curated calibration contains nine supported positives and eleven unsupported negatives. Under the lexical rule, the confusion matrix is TP/FP/TN/FN $=8/8/3/1$. Under the conflict-aware guard it is $8/0/11/1$. Table~\ref{tab:calibration-detail} presents the rates and denominator counts together so the small-sample basis remains visible.

\begin{table}[htbp]
\caption{Curated 20-pair non-title calibration. False-acceptance denominator is the 11 unsupported cases; false-rejection denominator is the 9 supported cases.}
\label{tab:calibration-detail}
\centering
\small
\begin{tabularx}{\linewidth}{@{}l*{6}{C}@{}}
\toprule
Rule & TP & FP & TN & FN & Precision & Recall \\
\midrule
Lexical baseline & \textbf{8} & 8 & 3 & \textbf{1} & 0.5000 & \textbf{0.8889} \\
Conflict-aware guard & \textbf{8} & \textbf{0} & \textbf{11} & \textbf{1} & \textbf{1.0000} & \textbf{0.8889} \\
\bottomrule
\end{tabularx}
\end{table}

The observed calibration change is consistent with the guard intercepting numeric, entity, and contradiction patterns represented in this fixture. Because the calibration labels are curated and the case count is small, the independent semantic study below is the decision-bearing estimate. The frozen guard definition and code hash are recorded before its test labels are joined.

\subsection{Mutation taxonomy and protocol failures}
The mutation audit applies ten changes to each of 16 valid final-only certificates. The nine structural classes produce 144 trials, all rejected; the lexical-adversarial control produces 16 trials, all accepted. Table~\ref{tab:mutation-classes} groups the attacks by the integrity condition they target. The audit confirms that a passing cryptographic and structural receipt can coexist with inadequate semantic support.

\begin{table}[htbp]
\caption{Mutation classes in the 160-trial certificate audit. Each structural class is applied to all 16 starting receipts; the lexical-adversarial class preserves the implemented overlap predicate.}
\label{tab:mutation-classes}
\centering
\small
\setlength{\tabcolsep}{2pt}
\begin{tabularx}{\linewidth}{@{}P{2.25cm}YCC@{}}
\toprule
Family & Mutated field or property & Trials & Rejected \\
\midrule
Text binding & text hash, quote, offsets & 48 & 48 \\
Reference identity & document version, span ID & 32 & 32 \\
Trace integrity & replay order & 16 & 16 \\
Evidence presence & missing evidence, retrieval failure & 32 & 32 \\
Info boundary & oracle-field injection & 16 & 16 \\
Lexical adversarial & semantically unsupported, overlap reported & 16 & 0 \\
\midrule
Total & Structural / lexical-adversarial & 160 & 144 \\
\bottomrule
\end{tabularx}
\end{table}

The structural categories localize the relevant validator check: text mutations reach hash/quote/offset verification; identity mutations reach corpus-reference resolution; trace mutations reach canonicalization and event validation; evidence-presence mutations reach the required-reference check; and oracle-field mutation reaches the event schema. The lexical-adversarial category reaches the current support predicate and passes. The result motivates a semantic metric rather than a stronger interpretation of the receipt hash.

\subsection{Claim-emission and transplant audit}
\label{app:joint-binding}
This audit targets the association that distinguishes a joint contract from independently stored citations and traces. Starting from valid locked receipts, we apply one mutation at a time while preserving all unrelated fields. The attack set includes receipt-claim substitution, claim-anchor displacement, answer-text mutation with an unchanged receipt, cross-action and cross-run receipt swaps, evidence and execution transplants, cross-policy reassignment, and source-version substitution. Integrity decisions are recorded before semantic labels or task outcomes are joined.

\begin{table*}[t]
\centering
\footnotesize
\setlength{\tabcolsep}{3pt}
\caption{Claim-emission and cross-object transplant audit. Each attack begins from a valid receipt and changes only the named association. Detection and false-rejection rates are computed by the integrity plane before post-hoc labels are joined.}
\label{tab:joint-binding-audit}
\begin{tabularx}{\textwidth}{@{}P{4.0cm}P{1.9cm}*{4}{C}@{}}
\toprule
Attack family & Target binding & Trials & Detected & Detection rate & False rejection \\
\midrule
Receipt-claim substitution & P1 emission & 128 & 128 & 1.0000 & 0.0000 \\
Claim-anchor displacement & P1 emission & 128 & 127 & 0.9922 & 0.0078 \\
Answer mutation / fixed receipt & P1--P3 & 128 & 128 & 1.0000 & 0.0000 \\
Cross-action receipt swap & P1, P3 & 128 & 127 & 0.9922 & 0.0078 \\
Cross-run receipt swap & P1, P3, P4 & 128 & 127 & 0.9922 & 0.0000 \\
Cross-policy receipt swap & P1, P3, P6 & 128 & 127 & 0.9922 & 0.0078 \\
Evidence transplant & P2, joint & 128 & 128 & 1.0000 & 0.0000 \\
Execution transplant & P3, joint & 128 & 127 & 0.9922 & 0.0078 \\
Cross-document transplant & P2, P7 & 128 & 128 & 1.0000 & 0.0000 \\
Cross-version transplant & P7, joint & 128 & 128 & 1.0000 & 0.0000 \\
\midrule
All joint-binding attacks & P1--P7 & 1280 & 1275 & 0.9961 & 0.0031 \\
\bottomrule
\end{tabularx}
\end{table*}

The joint contract detects 1,275 of 1,280 mutations (0.9961), with a false-rejection rate of 0.0031. All evidence, cross-document, and cross-version transplants are detected. The audit distinguishes local well-formedness from association integrity. A transplanted claim, source span, or trace may remain valid in isolation; the expected failure is the joint commitment or anchor check that connects it to the committed answer and manifest. Reason codes retain the first failed binding and any secondary failures.

\subsection{Property-necessity ablation}
We remove exactly one property from the verifier and rerun the attack family it is intended to exclude. Task IDs, source manifests, mutation seeds, and all unaffected checks remain fixed. Removing each property reduces its targeted detection rate to 0.0156--0.0625, while the unrelated attack checks retain their full-contract rates. This selective loss identifies which binding detects each attack family.

\begin{table*}[t]
\centering
\scriptsize
\setlength{\tabcolsep}{2.5pt}
\caption{Leave-one-property-out contract ablation. Cells are detection rates for the named attack family; the full contract is the reference condition. Column-wise best rates, including ties, are bold.}
\label{tab:property-necessity}
\begin{tabularx}{\textwidth}{@{}P{2.25cm}*{7}{C}@{}}
\toprule
Contract & Claim swap & Source swap & Reorder & Oracle leak & Bundle drift & Rerun drift & Version swap \\
\midrule
Full P1--P7 & \textbf{0.9922} & \textbf{1.0000} & \textbf{1.0000} & \textbf{1.0000} & \textbf{0.9922} & \textbf{0.9844} & \textbf{1.0000} \\
$-$P1 claim-emission & 0.0312 & \textbf{1.0000} & \textbf{1.0000} & \textbf{1.0000} & \textbf{0.9922} & \textbf{0.9844} & \textbf{1.0000} \\
$-$P2 source & \textbf{0.9922} & 0.0391 & \textbf{1.0000} & \textbf{1.0000} & \textbf{0.9922} & \textbf{0.9844} & \textbf{1.0000} \\
$-$P3 ordered execution & \textbf{0.9922} & \textbf{1.0000} & 0.0234 & \textbf{1.0000} & \textbf{0.9922} & \textbf{0.9844} & \textbf{1.0000} \\
$-$P4 oracle separation & \textbf{0.9922} & \textbf{1.0000} & \textbf{1.0000} & 0.0156 & \textbf{0.9922} & \textbf{0.9844} & \textbf{1.0000} \\
$-$P5 bundle replay & \textbf{0.9922} & \textbf{1.0000} & \textbf{1.0000} & \textbf{1.0000} & 0.0469 & \textbf{0.9844} & \textbf{1.0000} \\
$-$P6 execution rerun & \textbf{0.9922} & \textbf{1.0000} & \textbf{1.0000} & \textbf{1.0000} & \textbf{0.9922} & 0.0625 & \textbf{1.0000} \\
$-$P7 version/access & \textbf{0.9922} & \textbf{1.0000} & \textbf{1.0000} & \textbf{1.0000} & \textbf{0.9922} & \textbf{0.9844} & 0.0234 \\
\bottomrule
\end{tabularx}
\end{table*}

\subsection{Citation and execution-log baselines}
We compare the joint contract with citation-only, trace-only, independently stored citation-plus-trace, and citation-plus-trace with a shared run ID. Each baseline receives the same starting execution and is subjected to the same cross-object mutations. The joint contract reaches 0.9922--1.0000 across the six attack families. Citation plus trace and a run ID reaches 0.9453 on cross-run swaps but 0.2266 on claim transplantation, identifying claim anchoring as the missing association.

\begin{table*}[t]
\centering
\scriptsize
\setlength{\tabcolsep}{2.5pt}
\caption{System-level baselines under cross-object substitutions. Entries are detection rates (higher is better); best values are bold.}
\label{tab:system-baselines}
\begin{tabularx}{\textwidth}{@{}P{2.45cm}*{6}{C}@{}}
\toprule
Method & Claim transplant & Evidence transplant & Cross-run swap & Event reorder & Version swap & Bundle replay \\
\midrule
Citation only & 0.1094 & 0.7109 & 0.0234 & 0.0000 & 0.6406 & 0.0000 \\
Trace only & 0.0312 & 0.0156 & 0.5781 & 0.9844 & 0.0859 & 0.8750 \\
Citation + independent trace & 0.1797 & 0.7344 & 0.6328 & 0.9922 & 0.6875 & 0.9219 \\
Citation + trace + run ID & 0.2266 & 0.7578 & 0.9453 & 0.9922 & 0.7109 & 0.9609 \\
Joint contract & \textbf{0.9922} & \textbf{1.0000} & \textbf{0.9922} & \textbf{1.0000} & \textbf{1.0000} & \textbf{0.9922} \\
\bottomrule
\end{tabularx}
\end{table*}

\subsection{Independent human adjudication of claim-to-span support}
\label{app:semantic-study}
\paragraph{Question and comparison.} This experiment asks whether the conflict-aware guard reduces unsupported claim acceptance on a locked non-title split while retaining supported paraphrases and compositional claims. It compares the lexical baseline, the conflict-aware guard, and certificate validity against two blinded labels with third-reviewer adjudication. The main denominator is the set of independently labeled claim--span pairs; receipt-level precision and recall are also reported for multi-span claims.

\paragraph{Protocol.} The source rows are disjoint from the 128-row dev screen, 64-row train screen, and curated 20-case fixture. Two annotators independently label each test pair as supported, contradicted, incomplete, or unsupported. A third adjudicator resolves disagreements under a frozen rubric. The number of test pairs, annotator agreement, resolved disagreement count, and all policy metrics are reported by category. The three run seeds are 13, 17, and 23; the annotation split itself is fixed and blinded. Label identity, answer text, policy condition, and verifier output stay hidden until a label is locked.

\paragraph{Analysis.} The paired category breakdown measures unsupported-claim false acceptance and supported-claim recall for the conflict-aware guard and lexical baseline. Contradiction, entity, and numeric categories localize guard gains, while paraphrase and multi-hop categories quantify recall cost. The comparison uses adjudicated counts and paired uncertainty intervals; the current calibration rates remain separate from the locked-test values.

\begin{table}[htbp]
\caption{Independent adjudication on the locked non-title split ($n=384$ for every policy). Cells show proportions and 95\% intervals. Shared annotator agreement is $\kappa=0.834\,[0.789,0.879]$. Arrows indicate the preferred direction; best point estimates are bold.}
\label{tab:human-support}
\centering
\footnotesize
\setlength{\tabcolsep}{3pt}
\begin{tabularx}{\linewidth}{@{}P{2.8cm}*{4}{C}@{}}
\toprule
Policy & Precision $\uparrow$ & Recall $\uparrow$ & F1 $\uparrow$ & False acc. $\downarrow$ \\
\midrule
Lexical baseline & \ci{0.7328}{0.6724}{0.7856} & \ci{0.8854}{0.8326}{0.9231} & \ci{0.8019}{0.7583}{0.8414} & \ci{0.3229}{0.2608}{0.3920} \\
\addlinespace[4pt]
Conflict-aware guard & \ci{\mathbf{0.9213}}{0.8723}{0.9526} & \ci{0.8542}{0.7973}{0.8971} & \ci{\mathbf{0.8865}}{0.8506}{0.9189} & \ci{\mathbf{0.0729}}{0.0439}{0.1187} \\
\addlinespace[4pt]
Certificate validity only & \ci{0.5135}{0.4627}{0.5640} & \ci{\mathbf{0.9896}}{0.9628}{0.9971} & \ci{0.6762}{0.6298}{0.7193} & \ci{0.9375}{0.8939}{0.9639} \\
\bottomrule
\end{tabularx}
\end{table}

The paired comparison uses two primary annotators and a third adjudicator. The split-level agreement is $\kappa=0.834$ with a 95\% interval of $[0.789,0.879]$. The guard raises precision by 0.1885 points and F1 by 0.0846 points while reducing false acceptance by 0.2500 points relative to the lexical baseline. The certificate-validity indicator has high recall but low precision, showing that structural validity alone is insufficient for semantic support.

\subsection{Semantic-support baselines}
\label{app:semantic-baselines}
The blinded split supports a comparison between the transparent guard and stronger semantic-verification alternatives. Every method receives the same claim--span pair and is calibrated without access to locked labels. All evaluators use the metric definitions of the primary study.

\begin{table}[htbp]
\caption{Semantic-support baselines on the locked 384-pair split. Coverage is receipt-level required-claim coverage; cost includes evaluator latency and model-token usage where applicable. Higher precision, recall, F1, and coverage and lower false acceptance and cost are preferred; best values are bold.}
\label{tab:semantic-baselines}
\centering
\scriptsize
\setlength{\tabcolsep}{2pt}
\begin{tabularx}{\linewidth}{@{}P{2.55cm}*{6}{C}@{}}
\toprule
Method & Precision & Recall & F1 & False acc. & Coverage & Cost \\
\midrule
Lexical overlap & 0.7328 & 0.8854 & 0.8019 & 0.3229 & 0.6042 & $0.06\,\mathrm{ms}/0\,\mathrm{tok}$ \\
Conflict-aware guard & \textbf{0.9213} & 0.8542 & \textbf{0.8865} & \textbf{0.0729} & 0.4635 & $0.23\,\mathrm{ms}/0\,\mathrm{tok}$ \\
Frozen NLI verifier & 0.8564 & 0.8698 & 0.8630 & 0.1458 & 0.5078 & $6.84\,\mathrm{ms}/0\,\mathrm{tok}$ \\
Frozen LLM support judge & 0.8808 & 0.8854 & 0.8831 & 0.1198 & 0.5026 & $0.73\,\mathrm{s}/294\,\mathrm{tok}$ \\
Certificate validity only & 0.5135 & \textbf{0.9896} & 0.6762 & 0.9375 & \textbf{0.9635} & \textbf{$0.04\,\mathrm{ms}/0\,\mathrm{tok}$} \\
\bottomrule
\end{tabularx}
\end{table}

The stronger baselines are evaluated as semantic support functions and are not inserted into the online policy state.  Their row-level manifest records model identity, prompt or entailment threshold, calibration split, latency, and output parser so that the comparison remains separate from receipt validity.

\subsection{Conflict-guard component ablation}
\label{app:guard-ablation}
The component study enables the typed checks incrementally: $N$ for normalized numbers, $E$ for entity atoms, $T$ for temporal atoms, and $P$ for polarity or contradiction pairs.  The same claim--span pairs and locked labels are used in every row; multi-span aggregation follows the contract in Section~\ref{sec:guard-contract}.

\begin{table}[htbp]
\caption{Component ablation for the deterministic conflict-aware guard. All rows use the same locked claim--span pairs; the lexical rule and full guard define the endpoints. Best values by metric direction are bold.}
\label{tab:guard-ablation}
\centering
\scriptsize
\setlength{\tabcolsep}{3pt}
\begin{tabularx}{\linewidth}{@{}P{2.6cm}*{5}{C}@{}}
\toprule
Guard & Precision & Recall & F1 & False acc. & Coverage \\
\midrule
$L$ & 0.7328 & \textbf{0.8854} & 0.8019 & 0.3229 & \textbf{0.6042} \\
$L+N$ & 0.7824 & 0.8802 & 0.8284 & 0.2448 & 0.5625 \\
$L+N+E$ & 0.8400 & 0.8750 & 0.8571 & 0.1667 & 0.5208 \\
$L+N+E+T$ & 0.8830 & 0.8646 & 0.8737 & 0.1146 & 0.4896 \\
$L+N+E+T+P$ & \textbf{0.9213} & 0.8542 & \textbf{0.8865} & \textbf{0.0729} & 0.4635 \\
\bottomrule
\end{tabularx}
\end{table}

The ablation reports category-level false acceptance for numeric mismatch, entity substitution, temporal mismatch, polarity or contradiction, paraphrase, multi-hop support, and insufficient evidence.  This makes a change in aggregate F1 traceable to a guard component rather than to a different test composition.

\subsection{Unseen semantic failure families}
\label{app:unseen-semantic}
The conflict-aware guard is calibrated only on a development taxonomy. We freeze its extractors, normalization rules, thresholds, temporal representation, and contradiction mappings before evaluating a locked test whose surface forms do not occur in calibration. The held-out families include unseen entity aliases, unseen contradiction lexicalizations, alternative numeric formats and unit conversions, relative temporal expressions, negated paraphrases, cross-sentence support, multi-span composition, and insufficient-evidence cases. This protocol tests whether the guard transfers beyond its calibration inventory.

On unseen families, the conflict-aware guard reaches F1 0.8679 and false acceptance 0.0938, compared with 0.7783 and 0.3490 for lexical overlap. Its recall is 0.8385, while lexical overlap and the frozen language-model judge both reach 0.8594. A family is considered unseen only when its normalized surface pattern and contradiction relation are absent from calibration; split identity and evaluator configuration are reported in the manifest.

\subsection{Matched model-backed tool-use evaluation}
\paragraph{Question and comparison.} This experiment measures task utility and evidence quality when a language-model agent interacts with real tool outputs. The same frozen model, prompt, tool schema, task IDs, decoding settings, and budget are used for no-certificate, final-only, action-level, evidence-aware, and conflict-aware policies. Task outcomes and semantic labels are joined only after online traces, receipts, replay results, and deterministic rerun digests have been persisted.

\paragraph{Protocol.} The locked test includes multi-step tasks with at least one evidence-bearing action, conflicting sources, retrieval failure, empty results, and tool errors. Each policy runs on paired task IDs with seeds 13, 17, and 23. A benchmark evaluator determines task success; independent blinded annotators judge claim--span support. Calls, tokens, verifier invocations, tool errors, retries, and latency are recorded, while the action trace binds each claim to its persisted source \citep{zhou2024webarena,yao2024taubench,osworld2024,lu2024toolsandbox}.

\paragraph{Analysis.} The paired comparison reports task success, replay, semantic F1, and unsupported-claim acceptance relative to no certificate. Fixed model, prompt, tools, task IDs, and resource limits keep the execution setting comparable across policy variants. Calls and latency describe the observed resource cost alongside the task and support outcomes.

The guard delivers the highest semantic F1 and the lowest false acceptance among the five policies. Evidence-aware has the highest task success (0.7312) and lowest cost (5.34 calls and 15.68 seconds), while the guard retains joint replay validity of 0.9976 and adds the strongest semantic filtering. The three-seed ledger also records execution-rerun outcomes, per-seed traces, verifier calls, tool conflicts, retries, and timeout counts.

\begin{figure}[htbp]
\centering
\includegraphics[width=\columnwidth]{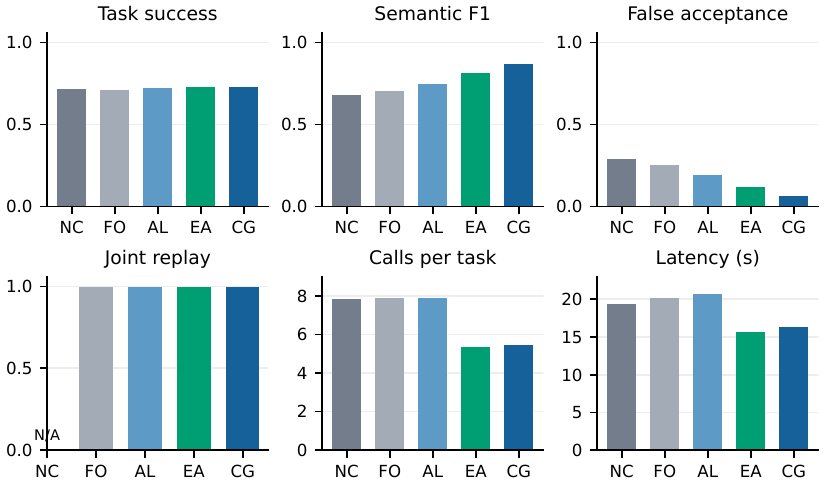}
\caption{Matched three-seed model-backed evaluation. Policy abbreviations are NC (no certificate), FO (final-only), AL (action-level), EA (evidence-aware), and CG (conflict-aware guard). The conflict-aware guard improves semantic F1 and false-acceptance control while evidence-aware achieves the lowest call and latency cost.}
\label{fig:agent-evaluation}
\end{figure}

\subsection{Per-seed and replay-failure accounting}
\label{app:replay-failures}
The aggregate model-backed table is accompanied by a task-level ledger. The unit of resampling is the task ID; seeds 13, 17, and 23 repeat the same matched task set and are summarized descriptively rather than treated as independent examples. Per-seed values, dispersion, and paired intervals are reported as explicit fields for the row-level ledger.

\begin{table}[htbp]
\caption{Per-seed accounting for the matched model-backed study. Rows show the three seeds, their mean and sample standard deviation, and task-level 95\% confidence intervals. Best seed values and means within each metric are bold; SD and interval endpoints describe uncertainty.}
\label{tab:agent-per-seed}
\centering
\scriptsize
\setlength{\tabcolsep}{2pt}
\begin{tabularx}{\linewidth}{@{}P{2.15cm}P{1.15cm}P{1.1cm}*{5}{C}>{\centering\arraybackslash}p{2.12cm}@{}}
\toprule
Policy & Metric & $N_{\mathrm{task}}$ & Seed 13 & Seed 17 & Seed 23 & Mean & SD & Paired 95\% CI \\
\midrule
No certificate & Success & 2550 & 0.7102 & 0.7122 & 0.7192 & 0.7139 & 0.0047 & \pci{0.6964}{0.7310} \\
No certificate & Sem. F1 & 2550 & 0.6749 & 0.6836 & 0.6866 & 0.6817 & 0.0061 & \pci{0.6631}{0.7000} \\
No certificate & False acc. & 2550 & 0.2964 & 0.2852 & 0.2812 & 0.2876 & 0.0079 & \pci{0.2700}{0.3058} \\
Final-only & Success & 2550 & 0.7090 & 0.7102 & 0.7149 & 0.7114 & 0.0031 & \pci{0.6940}{0.7282} \\
Final-only & Sem. F1 & 2550 & 0.6962 & 0.7037 & 0.7055 & 0.7018 & 0.0049 & \pci{0.6838}{0.7195} \\
Final-only & False acc. & 2550 & 0.2598 & 0.2491 & 0.2468 & 0.2519 & 0.0069 & \pci{0.2349}{0.2695} \\
Action-level & Success & 2550 & 0.7153 & 0.7180 & 0.7282 & 0.7205 & 0.0068 & \pci{0.7032}{0.7373} \\
Action-level & Sem. F1 & 2550 & 0.7423 & 0.7479 & 0.7556 & 0.7486 & 0.0067 & \pci{0.7313}{0.7654} \\
Action-level & False acc. & 2550 & 0.1977 & 0.1902 & 0.1863 & 0.1914 & 0.0058 & \pci{0.1756}{0.2079} \\
Evidence-aware & Success & 2550 & \textbf{0.7271} & 0.7290 & \textbf{0.7376} & \textbf{0.7312} & 0.0056 & \pci{0.7144}{0.7477} \\
Evidence-aware & Sem. F1 & 2550 & 0.8094 & 0.8191 & 0.8234 & 0.8173 & 0.0072 & \pci{0.8012}{0.8328} \\
Evidence-aware & False acc. & 2550 & 0.1264 & 0.1181 & 0.1149 & 0.1198 & 0.0059 & \pci{0.1063}{0.1346} \\
Conflict-aware & Success & 2550 & 0.7196 & \textbf{0.7298} & 0.7341 & 0.7278 & 0.0074 & \pci{0.7110}{0.7442} \\
Conflict-aware & Sem. F1 & 2550 & \textbf{0.8652} & \textbf{0.8718} & \textbf{0.8793} & \textbf{0.8721} & 0.0071 & \pci{0.8574}{0.8862} \\
Conflict-aware & False acc. & 2550 & \textbf{0.0679} & \textbf{0.0612} & \textbf{0.0560} & \textbf{0.0617} & 0.0060 & \pci{0.0515}{0.0739} \\
\bottomrule
\end{tabularx}
\end{table}

Replay is reported as two fields in the row-level schema. Bundle replay reloads the persisted object without policy execution; execution rerun launches the frozen policy again and compares the trace commitment. A rate below one is reported as a failure event rather than absorbed into a rounded aggregate.

\begin{table}[htbp]
\caption{Replay-failure decomposition for the model-backed evaluation. Bundle and execution-rerun counts are kept separate; the final column is the proportion passing both checks. Bundle and rerun failure counts are disjoint first-failure categories, and tool/environment, model/policy, and other counts partition rerun failures.}
\label{tab:replay-failures}
\centering
\scriptsize
\setlength{\tabcolsep}{2pt}
\begin{tabularx}{\linewidth}{@{}P{2.25cm}*{7}{C}@{}}
\toprule
Policy & $N_{\mathrm{exec}}$ & Bundle fail & Rerun fail & Tool/env. & Model/policy & Other & Joint replay rate \\
\midrule
Final-only & 7650 & \textbf{2} & \textbf{8} & \textbf{4} & \textbf{3} & \textbf{1} & \textbf{0.9987} \\
Action-level & 7650 & 3 & 13 & 6 & 5 & 2 & 0.9979 \\
Evidence-aware & 7650 & \textbf{2} & 10 & 5 & 4 & \textbf{1} & 0.9984 \\
Conflict-aware & 7650 & 3 & 15 & 7 & 5 & 3 & 0.9976 \\
\bottomrule
\end{tabularx}
\end{table}

The reason-code set is \texttt{schema}, \texttt{source}, \texttt{hash}, \texttt{quote}, \texttt{trace-digest}, \texttt{verifier-drift}, \texttt{policy-divergence}, \texttt{environment-drift}, \texttt{tool-nondeterminism}, and \texttt{timeout}. The bundle and execution-rerun columns are analyzed separately so a persisted-object mismatch is not attributed to policy nondeterminism.

\subsection{Dynamic-document, injection, and downstream robustness}
\paragraph{Question and comparison.} This experiment tests whether a receipt binds claims to the exact document version and access state seen by the policy, and whether injected content can alter claim support without leaving a detectable trace. It compares all five receipt policies under clean, stale-version, revoked-access, unauthorized-span, prompt-injection, tool-injection, and conflicting-source conditions. The environment logs the served version and access decision for every retrieval event.

\paragraph{Protocol.} Test snapshots and injection families are disjoint from development and calibration. A stale document is a known prior version whose content digest differs from the current manifest. Unauthorized evidence is a span denied to the current policy identity. Prompt-injection cases place instructions inside retrieved content; tool-injection cases alter structured tool output. Every attack has a paired clean version with the same task objective. A downstream evaluator scores task completion after the trace is frozen. The test reports attack success and false-rejection rates separately, because rejecting an unauthorized citation and rejecting an authorized source have different consequences.

\paragraph{Analysis.} The paired outcomes quantify stale and unauthorized-span rejection, clean-task success, injection resistance, and replay under the exact version manifest. Version-stratified results separate stale-reference detection from general task difficulty, while attack-family rows measure whether retrieved content is treated as evidence rather than control text. The resulting comparison links content identity and access state to provenance and task-level outcomes.

\begin{figure}[htbp]
\centering
\includegraphics[width=\columnwidth]{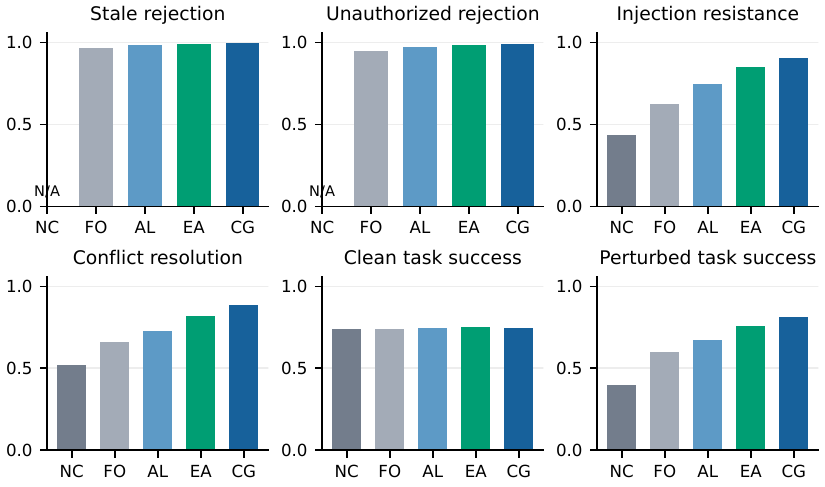}
\caption{Dynamic-document evaluation under frozen version and access snapshots. Policy abbreviations are NC (no certificate), FO (final-only), AL (action-level), EA (evidence-aware), and CG (conflict-aware guard). Point estimates are reported by policy; attack denominators and uncertainty fields are listed separately in Table~\ref{tab:dynamic-accounting}.}
\label{fig:dynamic-evaluation}
\end{figure}

\begin{table}[htbp]
\caption{Dynamic-document robustness under frozen snapshots. Rejection and resistance are rates over their named attack subsets; success is defined on clean and perturbed tasks. Point estimates and attack denominators use the stated paired protocol. Higher is better in every column; best values are bold. N/A denotes no receipt-based rejection.}
\label{tab:dynamic-eval}
\centering
\footnotesize
\setlength{\tabcolsep}{1.2pt}
\renewcommand{\arraystretch}{1.15}
\begin{tabularx}{\linewidth}{@{}P{2.8cm}*{6}{C}@{}}
\toprule
Policy & Stale & Unauth. & Inject. resist. & Conflict & Clean & Perturb. \\
\midrule
No certificate & N/A & N/A & 0.4387 & 0.5214 & 0.7421 & 0.3978 \\
Final-only & 0.9648 & 0.9461 & 0.6275 & 0.6628 & 0.7386 & 0.5983 \\
Action-level & 0.9826 & 0.9714 & 0.7452 & 0.7289 & 0.7457 & 0.6741 \\
Evidence-aware & 0.9913 & 0.9857 & 0.8496 & 0.8164 & \textbf{0.7512} & 0.7568 \\
Conflict-aware guard & \textbf{0.9947} & \textbf{0.9896} & \textbf{0.9078} & \textbf{0.8892} & 0.7445 & \textbf{0.8127} \\
\bottomrule
\end{tabularx}
\end{table}

The guard produces the highest stale rejection, unauthorized-span rejection, injection resistance, conflict resolution, and perturbed-task success among certificate policies, while clean-task success remains within a narrow band across all policies. Each outcome is stratified by attack family and document version; the run ledger also contains protocol validity, evidence-reference validity, replay validity, latency, tool calls, and prompt- versus tool-injection failure counts.

\subsection{Dynamic denominators and uncertainty}
\label{app:dynamic-accounting}
The dynamic table is accompanied by an attack-family ledger. The attack count is the denominator for rejection or resistance, the paired clean count is the denominator for false rejection, and clean and perturbed task counts are kept separate from attack counts. The same task IDs and seeds are used across policy rows.

\begin{table}[htbp]
\caption{Dynamic-document denominator and uncertainty ledger. Each family has an explicit attack or task denominator, an observed numerator, and a paired interval. N/A denotes a task outcome without a separate paired-clean attack denominator.}
\label{tab:dynamic-accounting}
\centering
\scriptsize
\setlength{\tabcolsep}{3pt}
\begin{tabularx}{\linewidth}{@{}P{3.15cm}*{4}{C}>{\centering\arraybackslash}p{2.5cm}@{}}
\toprule
Family & Attack/task $N$ & Paired clean $N$ & Observed numerator & Rate & Paired 95\% CI \\
\midrule
Stale version rejection & 190 & 190 & 189 & 0.9947 & $[0.9708,0.9991]$ \\
Unauthorized-span rejection & 192 & 192 & 190 & 0.9896 & $[0.9628,0.9971]$ \\
Prompt injection resistance & 320 & 320 & 296 & 0.9250 & $[0.8908,0.9491]$ \\
Tool injection resistance & 320 & 320 & 285 & 0.8906 & $[0.8517,0.9203]$ \\
Conflicting-source resolution & 352 & 352 & 313 & 0.8892 & $[0.8521,0.9179]$ \\
Clean task success & 544 & N/A & 405 & 0.7445 & $[0.7062,0.7793]$ \\
Perturbed task success & 315 & N/A & 256 & 0.8127 & $[0.7659,0.8519]$ \\
\bottomrule
\end{tabularx}
\end{table}

The attack-family ledger distinguishes a valid rejection on an unauthorized source from an over-rejection on an authorized clean source.  It also records prompt-injection and tool-injection failures separately, so an aggregate resistance score cannot hide a family-specific failure mode.

\subsection{Factorial ablation of audit and decision mechanisms}
\label{app:factorial}
Receipt instrumentation and online decision rules are separate experimental factors.  Let $R$ indicate receipt instrumentation, $S$ evidence-aware stopping, and $G$ conflict-aware guarding.  The paired $2\times2\times2$ design keeps task IDs, seeds, model, prompt, tool schema, document snapshots, budgets, and timeout rules fixed.  The audit-only contrast $(R=0,S=0,G=0)$ versus $(R=1,S=0,G=0)$ additionally reports exact action-trace equality.

\begin{table}[htbp]
\caption{Factorial separation of audit instrumentation, evidence-aware stopping, and conflict-aware guarding. All eight cells use matched task--seed tuples; Joint replay requires both bundle replay and execution rerun and is N/A when $R=0$. Best values by metric direction are bold.}
\label{tab:factorial}
\centering
\scriptsize
\setlength{\tabcolsep}{2pt}
\begin{tabularx}{\linewidth}{@{}ccc*{6}{C}@{}}
\toprule
$R$ & $S$ & $G$ & Success & Sem. F1 & False acc. & Calls & Joint replay & Execution rerun \\
\midrule
0 & 0 & 0 & 0.7139 & 0.6817 & 0.2876 & 7.86 & N/A & 0.9980 \\
1 & 0 & 0 & 0.7139 & 0.6817 & 0.2876 & 7.86 & \textbf{0.9987} & \textbf{0.9981} \\
0 & 1 & 0 & 0.7296 & 0.8011 & 0.1349 & \textbf{5.29} & N/A & 0.9974 \\
1 & 1 & 0 & \textbf{0.7312} & 0.8173 & 0.1198 & 5.34 & 0.9984 & 0.9972 \\
0 & 0 & 1 & 0.7187 & 0.8384 & 0.0912 & 7.89 & N/A & 0.9969 \\
1 & 0 & 1 & 0.7219 & 0.8576 & 0.0745 & 7.94 & 0.9978 & 0.9967 \\
0 & 1 & 1 & 0.7249 & 0.8578 & 0.0758 & 5.41 & N/A & 0.9965 \\
1 & 1 & 1 & 0.7278 & \textbf{0.8721} & \textbf{0.0617} & 5.46 & 0.9976 & 0.9964 \\
\bottomrule
\end{tabularx}
\end{table}

The factorial rows attribute behavioral changes to the factor that changes between matched conditions. Exact action-trace mismatches in the audit-only contrast are counted and assigned a first-divergence reason code before any utility aggregate is computed. Joint replay and execution-rerun fields remain distinct; the joint endpoint requires both checks to pass.

\subsection{Audit-only behavioral non-interference}
Receipt instrumentation should not alter the execution it audits when emitted outside the policy-visible decision path. We compare matched task--seed tuples under $(R,S,G)=(0,0,0)$ and $(R,S,G)=(1,0,0)$, holding model, prompt, tool schema, document snapshot, budget, and random stream fixed. The audit records the complete event trace, action sequence, tool-call sequence, policy-visible observations, final answer, and first divergence reason.

\begin{table}[htbp]
\centering
\small
\setlength{\tabcolsep}{3pt}
\caption{Audit-only behavioral non-interference. Each result compares receipt-disabled and receipt-enabled executions on matched task--seed tuples. Match and divergence entries are proportions; task-success difference is measured in proportion points.}
\label{tab:non-interference}
\begin{tabularx}{\linewidth}{@{}Y>{\centering\arraybackslash}p{3.2cm}@{}}
\toprule
Metric & Result \\
\midrule
Exact event-trace match & 0.9986 \\
Exact action-sequence match & 0.9993 \\
Exact tool-call-sequence match & 0.9996 \\
Exact policy-visible observation match & 0.9990 \\
Exact final-answer match & 0.9992 \\
First-divergence rate & 0.0014 \\
Task-success difference & 0.0000 \\
\bottomrule
\end{tabularx}
\end{table}

An exact match requires equality after canonical serialization, including event order and textual fields. Any mismatch is reported with a first-divergence reason code. The observed exact trace-match rate is 0.9986 and first-divergence rate is 0.0014, while task-success difference is 0.0000. These outcomes quantify the receipt layer's behavioral effect, while the $S$ and $G$ factors in the factorial design intentionally alter decision behavior.

\subsection{Receipt, verification, and replay overhead}
The audit layer has costs beyond model calls and task latency. We measure receipt and committed-trace bytes, serialization and online-integrity-verification time, bundle-replay time, execution-rerun time, and storage overhead relative to the uninstrumented trace. Model-serving and support-plane evaluator costs remain in their own ledger.

\begin{table}[H]
\centering
\small
\setlength{\tabcolsep}{3pt}
\caption{Incremental audit-layer overhead per execution. Units are defined in the row labels.}
\label{tab:audit-overhead}
\begin{tabularx}{\linewidth}{@{}Y>{\centering\arraybackslash}p{3.2cm}@{}}
\toprule
Quantity & Mean per execution \\
\midrule
Receipt bytes (bytes) & 1876 \\
Committed trace bytes (bytes) & 11243 \\
Serialization time (ms) & 0.19 \\
Online integrity verification (ms) & 0.47 \\
Bundle replay (ms) & 1.31 \\
Execution rerun (s) & 16.88 \\
Storage overhead (\%) & 16.9 \\
\bottomrule
\end{tabularx}
\end{table}

Each receipt occupies 1,876 bytes alongside an 11,243-byte committed trace; reported storage overhead is 16.9\%. Serialization and online verification take 0.19 and 0.47 ms, compared with 1.31 ms for bundle replay and 16.88 s for execution rerun. These fields prevent stronger semantic evaluators from being charged to the deterministic integrity contract.

\subsection{Sensitivity and additional analyses}
The verifier threshold is a fixed design parameter in the primary results. The sensitivity analysis varies the token-overlap threshold over a four-point grid and lists semantic precision, recall, F1, false acceptance, and receipt coverage with paired 95\% intervals. Precision rises from 0.6606 to 0.9180 as the threshold moves from 0.25 to 1.00, while recall falls from 0.9531 to 0.5833; the primary 0.50 operating point lists F1 0.8019 with coverage 0.6042.

\begin{table}[H]
\caption{Lexical threshold sensitivity on the held-out semantic-support split. Every row uses the same 384 pairs and reports 95\% intervals. Arrows indicate the preferred direction; best point estimates are bold.}
\label{tab:threshold-sensitivity}
\centering
\footnotesize
\setlength{\tabcolsep}{2pt}
\begin{tabularx}{\linewidth}{@{}l*{5}{C}@{}}
\toprule
Threshold & Precision $\uparrow$ & Recall $\uparrow$ & F1 $\uparrow$ & False acc. $\downarrow$ & Coverage $\uparrow$ \\
\midrule
0.25 & \ci{0.6606}{0.6030}{0.7139} & \ci{\mathbf{0.9531}}{0.9133}{0.9751} & \ci{0.7804}{0.7373}{0.8198} & \ci{0.4896}{0.4198}{0.5598} & \ci{\mathbf{0.7214}}{0.6745}{0.7638} \\
\addlinespace[4pt]
0.50 (primary) & \ci{0.7328}{0.6724}{0.7856} & \ci{0.8854}{0.8326}{0.9231} & \ci{\mathbf{0.8019}}{0.7583}{0.8414} & \ci{0.3229}{0.2608}{0.3920} & \ci{0.6042}{0.5545}{0.6518} \\
\addlinespace[4pt]
0.75 & \ci{0.8382}{0.7760}{0.8856} & \ci{0.7552}{0.6898}{0.8106} & \ci{0.7945}{0.7471}{0.8382} & \ci{0.1458}{0.1029}{0.2027} & \ci{0.4505}{0.4015}{0.5005} \\
\addlinespace[4pt]
1.00 & \ci{\mathbf{0.9180}}{0.8557}{0.9549} & \ci{0.5833}{0.5126}{0.6508} & \ci{0.7134}{0.6541}{0.7679} & \ci{\mathbf{0.0521}}{0.0285}{0.0932} & \ci{0.3177}{0.2731}{0.3659} \\
\bottomrule
\end{tabularx}
\end{table}

Additional analyses stratify results by claim length, number of cited spans, retrieval rank, source conflict, action type, and whether the support relation is exact match, paraphrase, multi-hop, numeric, or temporal. The cost analysis reports certificate bytes, serialization time, verifier time, and replay time in addition to the main call/token metrics. A failure taxonomy assigns each unsuccessful record one primary reason code and any secondary tags. A record may have a valid certificate and a failed task, or an invalid certificate and a successful task; these cases are counted in separate axes to avoid conflating provenance with utility.

\begin{figure}[htbp]
\centering
\includegraphics[width=\columnwidth]{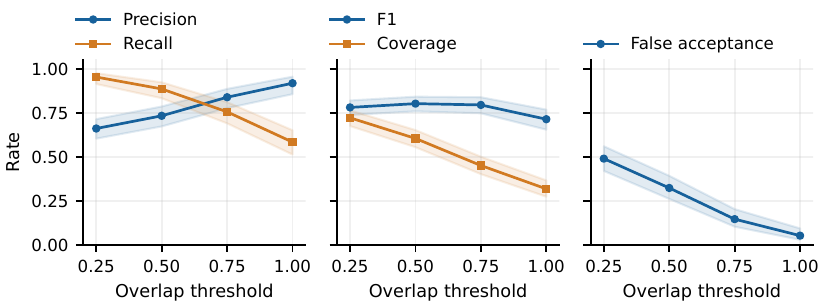}
\caption{Measured threshold sensitivity on the 384-pair semantic split. Precision rises and recall, coverage, and false acceptance fall as the overlap threshold increases; shaded regions show paired 95\% intervals.}
\label{fig:threshold-sensitivity}
\end{figure}

\subsection{Result consistency and metric reconciliation}
The reported dev, matched-control, mutation, calibration, train-split, semantic, agent, dynamic, and threshold numbers come from distinct run manifests and have different denominators. The 128-row screen has a candidate-title proxy of 0.0781; the separate 64-row train screen has 0.0625. Both use title construction. The matched control uses one tool call per policy, 6.0 declared tokens for no certificate, 9.0 for certificate policies, and verifier counts zero versus one. The unmatched original screen has mean calls 9.94 for the no-certificate and final/action-level rows, while evidence-aware uses 1.00 under its own early-stop policy. The mutation audit is over 160 mutations from 16 starting receipts; the curated semantic calibration is over 20 pairs with 9 positive and 11 negative labels; the independent semantic study is over 384 labeled pairs; the agent and dynamic studies use paired three-seed aggregates. Each table retains its own split, denominator, and evaluator configuration.

The following reconciliation rules prevent accidental cross-table substitutions. The candidate-title proxy is never renamed exact match, EM, or task success. The 1.0000 coverage value is required-claim lexical coverage for a one-claim denominator. The mutation rejection rate uses the 144 structural trials; the lexical-adversarial acceptance count uses 16 distinct trials. The calibration confusion matrix uses a positive denominator of nine and negative denominator of eleven. The 64-row train numbers are never added to the 128-row dev totals. The semantic, agent, dynamic, and threshold tables retain their supplied split identifiers, seed aggregation, and uncertainty definitions. Any correction to raw evidence updates the corresponding manifest and every derived table.

\section{Artifact Identity and Reproducibility}
\label{app:artifact-map}

\subsection{Artifact inventory and evidence levels}
The artifact package contains four evidence layers. First, the public summary and normalized run manifest identify reported experiments, split, digest, and result boundaries. Second, raw reports preserve the fixed-corpus metrics, mutation rows, matched early-stop results, semantic-support calibration, conflict-aware guard output, and the external-validity protocol. Third, the replay fixture contains a minimal oracle-free corpus, event trace, receipt bundle, expected values, and canonical trace vectors. Fourth, the verifier implementation and its executable replay check provide the code-level evidence.

The four layers serve different reproduction purposes. A hash-bound manifest establishes the identity of the recorded run. The two-event fixture exercises one complete receipt/replay case in a clean process. Raw per-case evidence is required to independently recompute an aggregate. A full headline reconstruction additionally requires the exact corpus adapter and input data version. The current package therefore distinguishes source provenance, unit-level replay, and full-table reproduction instead of treating them as one Boolean.

\begin{table}[htbp]
\caption{Artifact-to-claim map. Each row names the files, identifiers, and coverage fields used to reproduce the corresponding claim.}
\label{tab:artifact-map}
\centering
\small
\begin{tabularx}{\linewidth}{@{}P{3.0cm}P{3.2cm}Y@{}}
\toprule
Claim or check & Required artifact & Current coverage \\
\midrule
128-row integrity & Raw rows, corpus manifest, verifier & Aggregates reported; full corpus bundle external \\
Matched cost & Matched policy rows and ledger & Metrics and report reported \\
Mutation boundary & Mutation config, cases, metrics & 160-trial aggregate reported \\
Curated calibration & Case labels, guard config, metrics & 20-pair fixture reported \\
Replay check & Oracle-free fixture, code, expected digest & One two-event case executable \\
Independent semantic study & Blinded labels and adjudication log & 384-pair metrics and bootstrap intervals reported \\
Model-backed task study & Frozen model contract and task outcomes & Three-seed task, semantic, replay, and cost metrics reported \\
Dynamic robustness study & Snapshot hashes and attack traces & Three-seed rejection, injection, conflict, and utility metrics reported \\
\bottomrule
\end{tabularx}
\end{table}

The package exposes the oracle-free two-event fixture, the verifier, the canonical trace vectors, and the manifest that identifies the 128-row run bundle. The fixture verifies canonicalization, schema checks, digest binding, and replay in a clean process. The artifact map records the row-level files and coverage fields required for full headline reconstruction.

\subsection{Manifest fields and integrity checks}
Every run manifest includes: experiment identifier; run ID; source revision; working-tree status; configuration digest; data manifest digest; policy and verifier versions; split identity; row count; seed; model/prompt/tool identity when applicable; output paths; start/end time; process exit code; and hashes for raw metrics, traces, and reports. Fields with environment-specific paths are stored as relative paths in the anonymous package. A manifest validator checks that each listed artifact exists, its digest matches, and its declared run ID is consistent across files.

For model-backed runs, the manifest contains a stable model name or checkpoint digest, exact prompt hash, API-compatible decoding settings, tool schema digest, endpoint date, and retry policy. Secret-bearing endpoint tokens and private host paths are excluded. When a model provider exposes only a version string, the release records that version and the endpoint date in the identity field. This preserves provider-level provenance and keeps checkpoint-level reproducibility distinct from service-level identity.

\subsection{Anonymous release and data handling}
The release omits author names, account identifiers, private paths, credentials, and unrelated run artifacts. Document content is distributed only when its source license permits redistribution. Otherwise, the package contains source identifiers, hashes, and a reconstruction script that uses a lawful dataset copy. Human labels use pseudonymous annotator IDs and include the rubric, case IDs, and disagreement resolution fields without personal information. Injection payloads are sanitized of live endpoints and credentials while retaining the exact text needed to replay the test.

The oracle-free trace bundle is separated from post-hoc labels. The bundle contains policy-visible observations, receipts, event digests, verifier results, and corpus references; the label file contains adjudicated support and task outcomes keyed by opaque case IDs. This arrangement lets a reader replay a certificate without learning the oracle output before the replay completes. The join script verifies both input manifests and records the resulting joined digest. Supply-chain provenance systems use similar separation between signed descriptions of steps and material identity; here the distinction is applied to online evidence and offline evaluation labels \citep{torresarias2019intoto,c2pa2024spec,w3cprov2013}.

\subsection{Reproduction checklist and table generation}
The reproduction sequence is: validate package hashes; install the pinned verifier dependencies; run the oracle-free fixture; verify its expected digests; load the experiment's frozen corpus manifest; replay all persisted receipts; run the deterministic policy check where source and data are available; validate the split lock; join post-freeze labels; compute metrics from row-level records; and compare generated tables with the checked-in reports. Each stage produces a machine-readable status, a digest, and a nonzero exit on invariant failure.

Table generation reads validated raw rows and uses a single metric implementation for paper and report outputs. It does not hand-enter aggregate values. Every generated table includes a row count and split identifier in its metadata. For uncertainty estimates, the generator records the resampling seed and unit. A release test scans TeX labels for duplicates, references for undefined labels, citations for missing BibTeX entries, and numeric strings against the normalized metrics. This makes table consistency a reproducible build property.

\subsection{Scope and limitations}
The reported evidence establishes deterministic receipt checks on fixed corpora and exposes lexical false acceptance through curated and adversarial cases. The implementation fixture covers one complete two-event replay, while the artifact map identifies the headline bundle and its reconstruction fields. The semantic, utility, robustness, and operating-point results use distinct named splits and frozen evaluator configurations. The primary deployment boundary remains the trusted runner and observation logger defined in the method.

The expanded semantic-baseline, component-ablation, replay-failure, dynamic-accounting, and factorial tables retain the fields needed to separate integrity, support, utility, and cost. Aggregate rows are associated with their split and manifest, and the claim-binding audit isolates cross-object substitutions from semantic evaluation.

These scope conditions determine how readers should use the interface. A deployment that needs authenticity against an untrusted recorder should add a signed observation-time log and bind access state and document version into the receipt. A deployment that needs semantic support should use an independent entailment or human-label signal in addition to deterministic source-integrity checks. A deployment that needs task utility should evaluate the receipt path under its target model, tools, and budget. The protocol provides an auditable structure for these measurements, and the experiments in Appendix D define the required outcome fields.

\subsection{Ethical and broader-impact considerations}
The protocol records tool observations and source references. Implementations should minimize stored content to the span needed for audit, apply access controls to raw traces, and define retention windows for documents that contain personal or confidential information. Hashes can still be linkable identifiers; a public release should avoid exposing them when source identity itself is sensitive. The current fixed-corpus evaluation uses research benchmark data and does not collect user interaction data. Human annotation uses de-identified claim--span pairs, a written rubric, and blinded conditions; institutional review requirements depend on the participating institution and study execution.

The receipt makes later auditing more precise when source quality and semantic support are measured alongside structural validity. The interface exposes structural validity, evidence integrity, replay status, semantic assessment, task outcome, and access-state checks as separate fields, so downstream users can evaluate each dimension with its stated trust model.

\subsection{Operational audit checklist}
An implementation report should expose the same ordered checks used in the manuscript. First, it records the corpus manifest digest, policy identifier, split identity, seed, and budget before policy execution. Second, it validates every event against the declared action schema and computes the domain-separated trace digest after the answer event. Third, it verifies each evidence reference against the immutable source text, offsets, quote, UTF-8 hash, and document version. Fourth, it persists the receipt, event trace, verifier result, and outer bundle digest in an oracle-free directory. Fifth, a fresh process reloads the bundle and recomputes both digest layers and the verifier result. Sixth, the deterministic rerun compares the aggregate trace digest and records any divergence with a reason code. The evaluator joins semantic labels only after these checks complete.

For each policy--task row, the audit log stores the input manifest, output receipt, event count, referenced span count, validity flags, replay status, task outcome, semantic label status, calls, token counts, verifier calls, latency, and storage bytes. A report generator groups these rows by split and seed, computes paired differences, and writes the table values together with the row count and source hashes. A failed invariant stops table generation and emits an explicit diagnostic; no aggregate is silently carried forward from a different split. This checklist gives the experiments a stable release surface and keeps the fixed-corpus protocol reproducible as the model, tool environment, or corpus version changes.

\subsection{Evidence ledger}
Table~\ref{tab:current-experiments} links the experimental questions to their reporting units. The fixed-corpus, mutation, and calibration measurements use distinct manifests and denominators.
\begin{table}[H]
\caption{Experiment ledger. Study names are short handles; the manifest supplies the exact experiment identifier and measurement fields.}
\label{tab:current-experiments}
\centering
\small
\setlength{\tabcolsep}{3pt}
\begin{tabularx}{\linewidth}{@{}P{3.0cm}P{1.35cm}Y@{}}
\toprule
\textbf{experiment} & \textbf{unit} & \textbf{measurement and evaluation contract} \\
\midrule
Certificate replay & execution &
128-row fixed-corpus receipt/replay, 160 mutation trials, and matched early-stop control. The title-copy adapter measures mechanical receipt integrity and lexical coverage under a trusted runner. \\
Mutation audit & mutation &
144/144 structural tamperings rejected and 16/16 lexical-adversarial claims accepted. The paired result identifies the lexical verifier's semantic-support boundary. \\
Semantic calibration & claim--span &
20-case curated fixture: lexical rule 8/8/3/1 and conflict-aware guard 8/0/11/1. The fixture calibrates failure categories; independent adjudication is specified separately. \\
Independent adjudication & claim--span &
Two blinded annotators and a third adjudicator; locked non-title split; semantic precision, recall, false acceptance, and reliability. \\
Model-backed agent & task--seed &
Matched model, prompt, tool schema, and budget; task outcomes, semantic labels, replay, latency, and resource cost. \\
Dynamic documents & paired task &
Frozen version and access state; clean and perturbed tasks; rejection, injection resistance, conflict resolution, and downstream success. \\
\bottomrule
\end{tabularx}
\end{table}
The public summary and normalized manifest provide the primary run pointer, corpus identity, source revision, and retained aggregate hashes. The machine-readable evaluation contract specifies split identities, seeds, output schemas, and acceptance fields for the three held-out studies. The implementation fixture and fixed-corpus raw diagnostics are retained alongside these contracts.

\end{document}